\documentclass{beatcs}

\usepackage{amsfonts,amsthm,amssymb,amsmath}

\newcommand{\e}{\mathrm{e}}
\newcommand{\argmax}{\mathrm{argmax}}
\newcommand{\argmin}{\mathrm{argmin}}
\renewcommand{\mid}{\,|\,}
\newcommand{\pin}{p_\mathrm{in}}
\newcommand{\pout}{p_\mathrm{out}}
\newcommand{\cin}{c_\mathrm{in}}
\newcommand{\cout}{c_\mathrm{out}}
\newcommand{\id}{\mathbf{1}}
\newcommand{\ER}{Erd\H{o}s-R\'enyi}
\newcommand{\tr}{\mathrm{tr}}
\newcommand{\contig}{\trianglelefteq}
\newcommand{\Exp}{\mathbb{E}}
\newcommand{\eps}{\epsilon}

\begin{document}

\title{The Computer Science and Physics of Community Detection: Landscapes, Phase Transitions,\\and Hardness}
\author{Cristopher Moore\\
 Santa Fe Institute\\
 \texttt{moore@santafe.edu}}
\maketitle

\begin{abstract}
Community detection in graphs is the problem of finding groups of vertices which are more densely connected than they are to the rest of the graph.  This problem has a long history, but it is undergoing a resurgence of interest due to the need to analyze social and biological networks.  While there are many ways to formalize it, one of the most popular is as an inference problem, where there is a ``ground truth'' community structure built into the graph somehow.  The task is then to recover the ground truth knowing only the graph.  

Recently it was discovered, first heuristically in physics and then rigorously in probability and computer science, that this problem has a \emph{phase transition} at which it suddenly becomes impossible.  Namely, if the graph is too sparse, or the probabilistic process that generates it is too noisy, then no algorithm can find a partition that is correlated with the planted one---or even tell if there are communities, i.e., distinguish the graph from a purely random one with high probability.  Above this information-theoretic threshold, there is a second threshold beyond which polynomial-time algorithms are known to succeed; in between, there is a regime in which community detection is possible, but conjectured to require exponential time.  

For computer scientists, this field offers a wealth of new ideas and open questions, with connections to probability and combinatorics, message-passing algorithms, and random matrix theory.  Perhaps more importantly, it provides a window into the cultures of statistical physics and statistical inference, and how those cultures think about distributions of instances, landscapes of solutions, and hardness.
\end{abstract}

\section{Introduction}

A friend knocks on your office door, having made the journey across campus from the sociology department.  ``In my fieldwork, I found a network of social links between individuals,'' they say.  ``I think there are two demographic groups of equal size, and that people tend to link mostly to others within the same group.  But their demographic information is hidden from me.  If I show you the links, can you figure out where the two groups are?''

``I know this problem!'' you say.  ``We call it \textsc{Min Bisection}.  The question is how to divide the vertices into two equal groups, with as few edges as possible crossing from one group to the other.  It's NP-hard~\cite{garey-johnson}, so we can't expect to solve it perfectly in polynomial time, but there are lots of good heuristics.  Let me take a look at your graph.''  

\begin{figure}
\begin{center}
\includegraphics[width=4in]{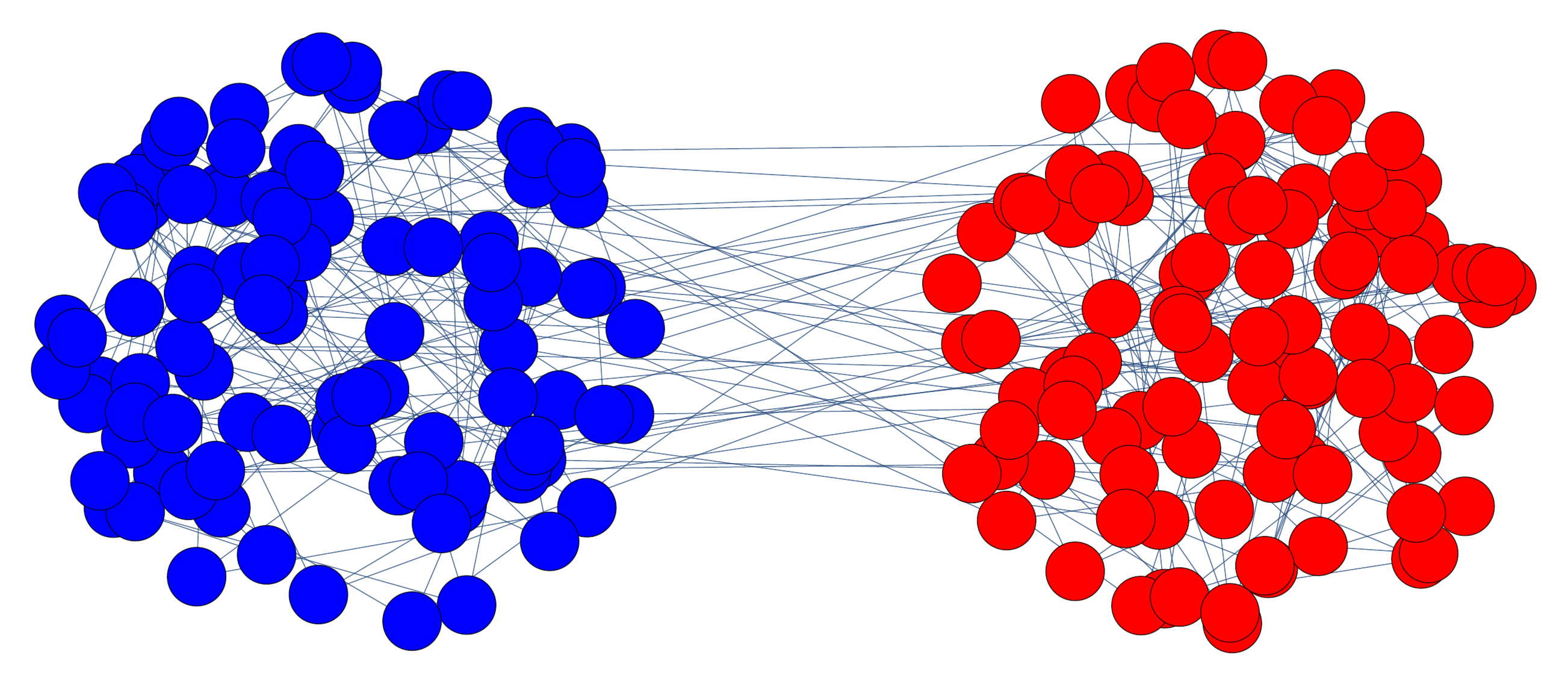}
\includegraphics[width=4in]{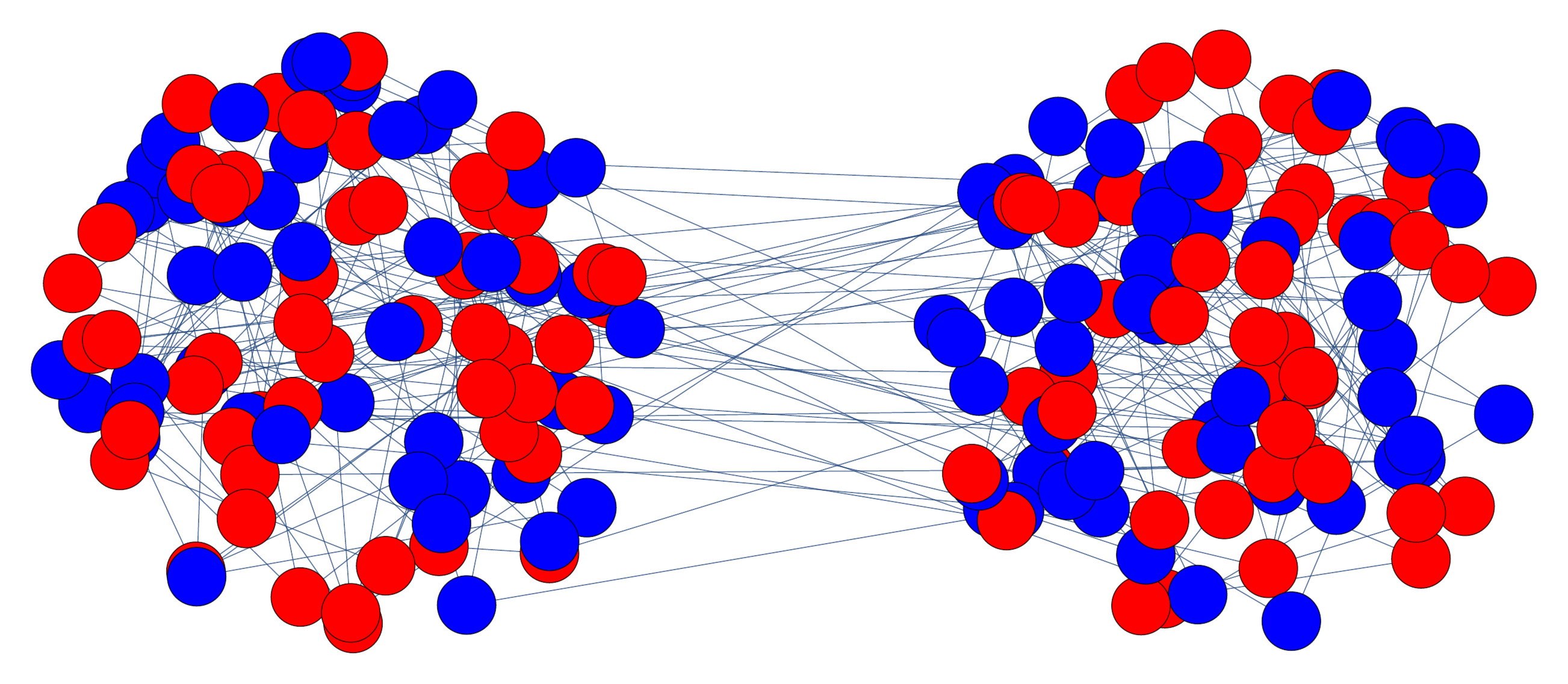}
\end{center}
\caption{Top: a partition of a graph into two equal groups, with $38$ of the $300$ edges crossing the cut.  Bottom: another partition of the same graph, with only $36$ edges crossing.  We kept the colors the same, and moved the vertices.  The two partitions are nearly uncorrelated: about half of each group in the first partition ended up on each side of the second one.  This is a sign that we might be overfitting, and in fact these ``communities'' exist only by chance: this is a random 3-regular graph.  Calculations from physics~\cite{zdeborova-boettcher} suggest that such graphs typically have bisections with only about $11\%$ of their edges crossing the cut; see also~\cite{dembo-montanari-sen}.  For evidence that this kind of overfitting and ambiguity can occur in real-world contexts as well, see~\cite{good-montjoye-clauset}.\label{fig:3reg}}
\end{figure}

You quickly code up a local search algorithm and find a local optimum.  You show it to your friend the sociologist, coloring the nodes red and blue, and putting them on the left and right side of the screen as in the top of Figure~\ref{fig:3reg}.  You proudly announce that of the $300$ edges in the graph, only $38$ of them cross between the groups.  It seems you have found a convincing division of the vertices (i.e., the people) into two communities.

``That's great!'' your friend says.  ``Is that the best solution?''

``Well, I just ran a simple local search,'' you say modestly.  ``We could try running it again, with different initial conditions\ldots''  You do so, and find a slightly better partition, with only $36$ edges crossing the cut.  ``Even better!'' says your friend.

But when you compare these two partitions (see the bottom of Figure~\ref{fig:3reg}, where we kept the colors the same but moved the vertices) you notice something troubling.  There is almost no correlation between them: they disagree about which person belongs in which group about as often as a coin flip.  ``Which one is right?'' your friend wants to know.

``Well, this is the better solution, so if you want the one with the smallest number of edges between groups, you should take the second one,'' you say.  ``\textsc{Min Bisection} is an optimization problem, inspired by assigning tasks to processors, or designing the layout of a circuit.  If you want to minimize the amount of communication or wiring you need, you should take whichever solution has the fewest edges between the groups.''

Your friend frowns.  ``But I'm not trying to design something---I'm trying to find out the truth.  If there are two ways to fit a model to my data that are almost as good as each other, but that don't agree at all, what am I supposed to think?  Besides, my data is uncertain---people might not report all their friendships, or they might be unusually willing to link with the other group.  A few small errors could flip which of these two solutions is better!''

At that moment, a physicist strides into the room.  ``Hi!'' she says brightly.  ``I couldn't help overhearing you.  It sounds to me like you have a bumpy energy landscape, with lots of local optima that are far apart from each other.  Even if you could find the ground state, it's not representative of the Boltzmann distribution at thermal equilibrium.  Besides, if it's hidden behind a free energy barrier, you'll never find it---you'll be stuck in a metastable paramagnetic state.  We see this in spin glasses all the time!''

You and your friend absorb this barrage of jargon and blink a few times.  ``But what should I do?'' asks the sociologist.  ``I want to label the individuals according to the two communities, and the only data I have is this graph.  How can I find the right way to do that?''

A stately statistician with a shock of gray in his hair walks by.  ``Perhaps there are in fact no communities at all!'' he declaims.  ``Have you solved the hypothesis testing problem, and decisively rejected the null?  For all you know, there is no real structure in your data, and these so-called `solutions' exist purely by chance.''  He points a shaking finger at you.  ``I find you guilty of overfitting!'' he thunders.

A combinatorialist peeks around the corner.  ``As a matter of fact, even random graphs have surprisingly good bisections!'' she says eagerly.  ``Allow me to compute the first and second moments of the number of bisections of a random graph with a fraction $\alpha$ of their edges crossing the cut, and bound the large deviation rate\ldots''  She seizes your chalk and begins filling your blackboard with asymptotic calculations.

At that moment, a flock of machine learning students rappel down the outside of the building, and burst in through your window shouting acronyms.  ``We'll use an EM algorithm to find the MLE!'' one says.  ``No, the MAP!'' says another.  ```Cross-validate!'' says a third.  ``Compute the AIC, BIC, AUC, and MDL!''  The physicist explains to anyone who will listen that this is merely a planted Potts model, an even more distinguished statistician starts a fistfight with the first one, and the sociologist flees.  Voices rise and cultures clash\ldots

\newpage

\section{The Stochastic Block Model}

There are many ways to think about community detection.  A natural one is to optimize some objective function, like \textsc{Min Bisection}.  One of the most popular such functions is the modularity~\cite{newman-girvan}, which measures how many edges lie within communities compared to what we would expect if the graph were randomly rewired while preserving the vertex degrees (i.e., if the graph were generated randomly by the configuration model~\cite{bollobas-book}).  It performs well on real benchmarks, and has the advantage of not fixing the number of groups in advance.  While maximizing it is NP-hard~\cite{brandes-etal}, it is amenable to spectral methods~\cite{Newman06c} and semidefinite and linear programming relaxations~\cite{agarwal-kempe}.  However, just as the random graph in Figure~\ref{fig:3reg} has surprisingly good bisections, even random graphs often have high modularity~\cite{guimera-random,zhang-moore-pnas}, forcing us to ask when the resulting communities are statistically significant.\footnote{This is a good time to warn the reader that I will not attempt to survey the vast literature on community detection.  One survey from the physics point of view is~\cite{fortunato-survey}; a computer science survey complementary to this one is~\cite{abbe-survey}.}

I will take a different point of view.  I will assume that the graph is generated by a probabilistic model which has community structure built into it, and that our goal is to recover this structure from the graph.  This breaks with tradition in theoretical computer science.  For the most part we are used to thinking about worst-case instances rather than random ones, since we want algorithms that are guaranteed to work on any instance.  But why should we expect a community detection algorithm to work, or care about its results, unless there really are communities in the first place?  And when Nature adds noise to a data set, isn't it fair to assume that this noise is random, rather than diabolically designed by an adversary?  

Without further ado, let us describe the stochastic block model, which was first proposed in the sociology literature~\cite{HLL83}.  It was reinvented in physics and mathematics as the ``inhomogeneous random graph''~\cite{Soderberg02,BJR07} and in computer science as the planted partition problem (e.g.~\cite{jerrum-sorkin,condon-karp,mcsherry}).  As a model of real networks is quite naive, but it can be elaborated to include arbitrary degree sequences~\cite{dcsbm}, overlapping communities~\cite{ABFX08,ball-karrer-newman}, and so on.  

There are $n$ vertices which belong to $q$ groups.  Each vertex $i$ belongs to a group $\sigma_i \in \{1,\ldots,q\}$; we call $\sigma$ the planted or ``ground truth'' assignment.  Given $\sigma$, the edges are generated independently, where each pair of vertices is connected with a probability that depends only on their groups:
\begin{equation}
\label{eq:sbm-general}
\Pr[(i,j) \in E] = p_{\sigma_i,\sigma_j} \, ,
\end{equation}
where $p$ is some $q \times q$ matrix.  We will focus on the symmetric case where the $\sigma_i$ are chosen a priori independently and uniformly, and where $\Pr[(i,j) \in E]$ depends only on whether $i$ and $j$ are in the same or different groups, 
\begin{equation}
\label{eq:symmetric}
\Pr[(i,j) \in E] = \begin{cases} 
\pin & \mbox{if $\sigma_i=\sigma_j$} \\
\pout & \mbox{if $\sigma_i \ne \sigma_j$} \, .
\end{cases}
\end{equation}

We often assume that $\pin > \pout$, i.e., that vertices are more likely to connect to others in the same group, which is called \emph{homophily} or \emph{assortativity}.  But the disassortative case $\pin < \pout$ is also interesting.  In particular, the case $\pin=0$ corresponds to the planted coloring problem, where edges are only allowed between vertices of different colors.  

Intuitively, communities are harder to find when the graph is sparse, since we have less information about each vertex.  We will spend most of our time in the regime where 
\begin{equation}
\label{eq:sparse}
\pin = \frac{\cin}{n} \quad \text{and} \quad \pout = \frac{\cout}{n}
\end{equation}
for some constants $\cin, \cout$.  Since the expected size of each group is $n/q$, each vertex has in expectation $\cin/q$ neighbors in its own group and $\cout/q$ in each of the other groups.  Thus the expected degree of each vertex is
\begin{equation}
\label{eq:avgdeg}
c = \frac{\cin + (q-1)\cout}{q} \, ,
\end{equation}
and the degree distribution is asymptotically Poisson.  The case $\cin = \cout$ is the classic \ER\ random graph where every pair of vertices is connected with the same probability $c/n$, which we denote $G(n,p=c/n)$.  The constant-degree regime $c=O(1)$ is where many classic phase transitions in random graphs occur, such as the emergence of the giant component~\cite{bollobas-book} or the $k$-core~\cite{pittel-spencer-wormald}, as well as the threshold for $q$-colorability (e.g.~\cite{achlioptas-naor}).

Now, given the graph $G$, can you reconstruct the planted assignment?  (Due to symmetry, we only ask you to do this up to a permutation of the $q$ groups.)  For that matter, can you confirm that $G$ was generated by this model, and not a simpler one without community structure?  This gives rise to several possible tasks, which we would like to solve with high probability---that is, with probability over $G$ that tends to $1$ as $n \to \infty$.  

\emph{Exact reconstruction} consists of finding the planted assignment exactly, labeling every vertex correctly up to a permutation of the groups.  This is impossible unless the graph is connected, and (in the assortative case) if every vertex has a plurality of its neighbors in its own group.  Neither of these local properties hold with high probability unless the average degree grows as $\Omega(\log n)$, suggesting the parametrization $\pin=\cin (\log n)/n$ and $\pout=\cout (\log n)/n$.  When $\cin$ and $\cout$ are large enough and different enough, exact reconstruction is in fact possible~\cite{BC09}.  Recent work has determined the threshold condition precisely, along with polynomial-time algorithms that achieve it~\cite{abbe-bandeira-hall,abbe-sandon,HajekWuXuSDP14,HajekWuXuSDP15,agarwal-etal,mns-consistency}.  

\emph{Reconstruction} (sometimes called weak reconstruction or weak recovery) consists of labeling the vertices with an assignment $\tau$ which is correlated with the planted assignment $\sigma$.  There are several reasonable definitions of correlation or accuracy: we will use the maximum, over all $q!$ permutations $\pi$ of the groups, of the fraction of vertices $i$ such that $\tau_i = \pi(\sigma_i)$.  We want this fraction to be bounded above $1/q$, i.e., noticeably better than chance.  

Finally, \emph{Detection} is the hypothesis testing problem that the statistician wanted us to solve: can we distinguish $G$ from an  \ER\ random graph $G(n,c/n)$ with the same average degree?  Intuitively, if we can't even tell if communities exist, we have no hope of reconstructing them.

I will assume throughout that the number of groups $q$ and the parameters $\cin$ and $\cout$ are known.  There are ways to learn $\cin$ and $\cout$ from the graph, at least when detection is possible; choosing $q$ is a classic ``model selection'' problem, and there are deep philosophical debates about how to solve it.  

I will focus on reconstruction and detection, both because that's what I know best, and because they have the strongest analogies with physics.  In particular, they possess phase transitions in the constant-degree regime.  If the community structure is too weak or the graph too sparse, they suddenly become impossible: no algorithm can label the vertices better than chance, or distinguish $G$ from a purely random graph.  

For $q \ge 5$ groups, and the disassortative case with $q = 4$, there are two distinct thresholds: a computational one above which efficient algorithms are known to exist, and an information-theoretic one below which the graph simply doesn't contain enough information to solve the problem.  In between, there is a regime where detection and weak reconstruction are possible, but believed to be exponentially hard---there is enough information in the graph to succeed, but we conjecture that any algorithm takes exponential time.  

In order to locate these phase transitions, we need to explore an analogy between statistical inference and statistical physics.

\section{Bayes and Ising}

It is easy to write down the probability $P(G \mid \sigma)$ that the block model will produce a given graph $G$ assuming a fixed group assignment $\sigma$.  Since the edges are generated independently according to~\eqref{eq:sbm-general}, this is just a product:
\[
P(G \mid \sigma) = \prod_{(i,j) \in E} p_{\sigma_i,\sigma_j} 
\prod_{(i,j) \notin E} \left(1 - p_{\sigma_i,\sigma_j} \right) \, .
\]
In the symmetric case~\eqref{eq:symmetric}, we can rewrite this as 
\begin{equation}
\label{eq:pgsigma}
P(G \mid \sigma) = 
\pout^m (1-\pout)^{{n \choose 2}-m} 
\prod_{(i,j) \in E} \left( \frac{\pin}{\pout} \right)^{\delta_{\sigma_i,\sigma_j}}
\prod_{(i,j) \notin E} \left( \frac{1-\pin}{1-\pout} \right)^{\delta_{\sigma_i,\sigma_j}}
\end{equation}
where $m$ is the total number of edges, and $\delta_{rs} = 1$ if $r=s$ and $0$ otherwise.  

Our goal is to invert the block model, recovering $\sigma$ from $G$.  In terms of Bayesian inference, this means computing the \emph{posterior} distribution: the conditional distribution of $\sigma$ given $G$, i.e., given the fact that the block model produced $G$.  According to Bayes' rule, this is
\begin{equation}
\label{eq:z}
P(\sigma \mid G) = \frac{1}{Z} \,P(G \mid \sigma) P(\sigma) 
\quad \text{where} \quad
Z = P(G) = \sum_{\sigma'} P(G \mid \sigma') P(\sigma') \, . 
\end{equation}
The normalization factor $Z$ is the total probability that the block model produces $G$, summed over all $q^n$ group assignments.  It is an important quantity in itself, but for now we simply note that it is a function only of $G$ and the parameters $\pin, \pout$ of the model, not of $\sigma$.  Furthermore, since we assumed that $\sigma$ is uniformly random, the prior probability $P(\sigma)$ is just the constant $q^{-n}$.  This leaves us with a simple proportionality
\[
P(\sigma \mid G) \propto P(G \mid \sigma) \, .
\]
We can also remove the constant before the products in~\eqref{eq:pgsigma}, giving
\begin{equation}
\label{eq:posterior}
P(\sigma \mid G) \propto \prod_{(i,j) \in E} \left( \frac{\pin}{\pout} \right)^{\delta_{\sigma_i,\sigma_j}}
\prod_{(i,j) \notin E} \left( \frac{1-\pin}{1-\pout} \right)^{\delta_{\sigma_i,\sigma_j}} \, . 
\end{equation}

In the assortative case $\pin > \pout$, this says that each edge $(i,j)$ of $G$ makes it more likely that $i$ and $j$ belong to the same group---since this edge is more likely to exist when that is true, and this edge does indeed exist.  Similarly, each non-edge $(i,j)$ makes it more likely that $i$ and $j$ belong to different groups.  In the sparse case where $\pin, \pout = O(1/n)$ the effect of the non-edges is weak, and we will often ignore them for simplicity, but they prevent too many vertices from being in the same group.

Many would call~\eqref{eq:posterior} a graphical model or Markov random field.  Each edge has a weight that depends on the state of its endpoints, and the probability of a state $\sigma$ is proportional to the product of these weights.  But it is also familiar to physicists, who like to describe probability distributions in terms of a \emph{Hamiltonian}\footnote{Yes, the same Hamilton the paths are named after.} or energy function $H(\sigma)$.  

Physical systems tend to be in low-energy states---rocks fall down---but thermal fluctuations kick them up to higher-energy states some of the time.  Boltzmann taught us that at equilibrium, the resulting probability distribution at equilibrium is
\begin{equation}
\label{eq:boltzmann}
P(\sigma) \propto \e^{-H(\sigma) / T} \, ,
\end{equation}
where $T$ is the temperature.  When the system is very hot, this distribution becomes uniform; as $T$ approaches absolute zero, it becomes concentrated at the ground states, i.e., the $\sigma$ with the lowest possible energy.  This we can think of $T$ as the amount of noise.

Physicists use the Hamiltonian to describe interactions between variables.  In a block of iron, each atom's magnetic field can be pointed up or down.  Neighboring atoms prefer to be aligned, giving them a lower energy if they agree.  Given a graph, the energy of a state $\sigma$ is
\begin{equation}
\label{eq:potts}
H(\sigma) = - J \sum_{(i,j) \in E} \delta_{\sigma_i,\sigma_j} \, ,
\end{equation}
for a constant $J$ which measures the strength of the interaction.  When $q=2$ this is the Ising model of magnetism, and for $q > 2$ it is called the Potts model.  The cases $J > 0$ and $J < 0$ are called \emph{ferromagnetic} and \emph{antiferromagnetic} respectively; iron is a ferromagnet, but there are also antiferromagnetic materials.

As the reader may already have divined, if we make the right choice of $J$ and $T$, the posterior distribution $P(\sigma \mid G)$ in the block model will be exactly the Boltzmann distribution of the Ising model on $G$.  If we compare~\eqref{eq:boltzmann} with our posterior distribution~\eqref{eq:posterior} and ignore the non-edges, we see that
\[
\frac{\pin}{\pout} = \e^{J/T} 
\quad \text{so} \quad 
\frac{J}{T} = \log \frac{\pin}{\pout} \, ,
\]
so assortativity and disassortativity correspond to ferromagnetic and antiferromagnetic models respectively.  (In the assortative case, the non-edges cause a weak global antiferromagnetic interaction, which keeps the groups of roughly equal size.)  As the community structure gets stronger, increasing $\pin/\pout$, we can think of this either as strengthening the interactions between neighbors, i.e., making $J$ bigger, or reducing the noise, i.e., making $T$ smaller.  

Of course, most magnets don't have a ``correct'' state.  The block model is a special case: there is a true underlying state $\sigma$, and $G$ is generated in a way that is correlated with $\sigma$.  We say that $G$ has $\sigma$ ``planted'' inside it.  
We can think of the block model as a noisy communication channel, where Nature is trying to tell us $\sigma$, but can only do so by sending us $G$.  The posterior distribution $P(\sigma \mid G)$ is all we can ever learn about $\sigma$ from $G$.  Even if we have unlimited computational resources, if this distribution doesn't contain enough information to reconstruct $\sigma$, there is nothing we can do.  The question is thus to what extent $P(\sigma \mid G)$ reveals the ground truth. 

In particular, suppose we can somehow find the ground state, i.e., the $\hat{\sigma}$ with the lowest energy and therefore the highest probability:
\[
\hat{\sigma} 
= \argmin_\sigma \,H(\sigma)
= \argmax_\sigma \,P(\sigma \mid G) \, . 
\]
Some people call $\hat{\sigma}$ the maximum likelihood estimate (MLE), or if we take the prior $P(\sigma)$ into account, the maximum a posteriori estimate (MAP).  Exact reconstruction is information-theoretically possible if and only if $\hat{\sigma} = \sigma$: that is, if the community structure is strong enough, or the graph is dense enough, that the most likely state is in fact the planted one.  As discussed above, this occurs when the average degree is a large enough constant times $\log n$.

But from an inference point of view, this seems somewhat optimistic.  Modern data sets are massive, but so is the number of variables we are typically trying to infer.  Getting all of them exactly right (in this case, all $n$ of the $\sigma_i$) seems to be asking a lot.  It is also strange from a physical point of view: it would be as if all $n=10^{26}$ atoms in a block of iron were aligned.  Physical systems are sparse---each atom interacts with just a few neighbors, and thermal noise will always flip a constant fraction of atoms the wrong way.  The most we can hope for is that the fraction of atoms pointing up, say, is some constant greater than $1/2$, so that the block of iron is magnetized---analogous to weak reconstruction, where the fraction of nodes labeled correctly is a constant greater than $1/q$.  Are there regimes where even this is too much to ask?

\begin{figure}
\begin{center}
\includegraphics[width=0.33\columnwidth,angle=90]{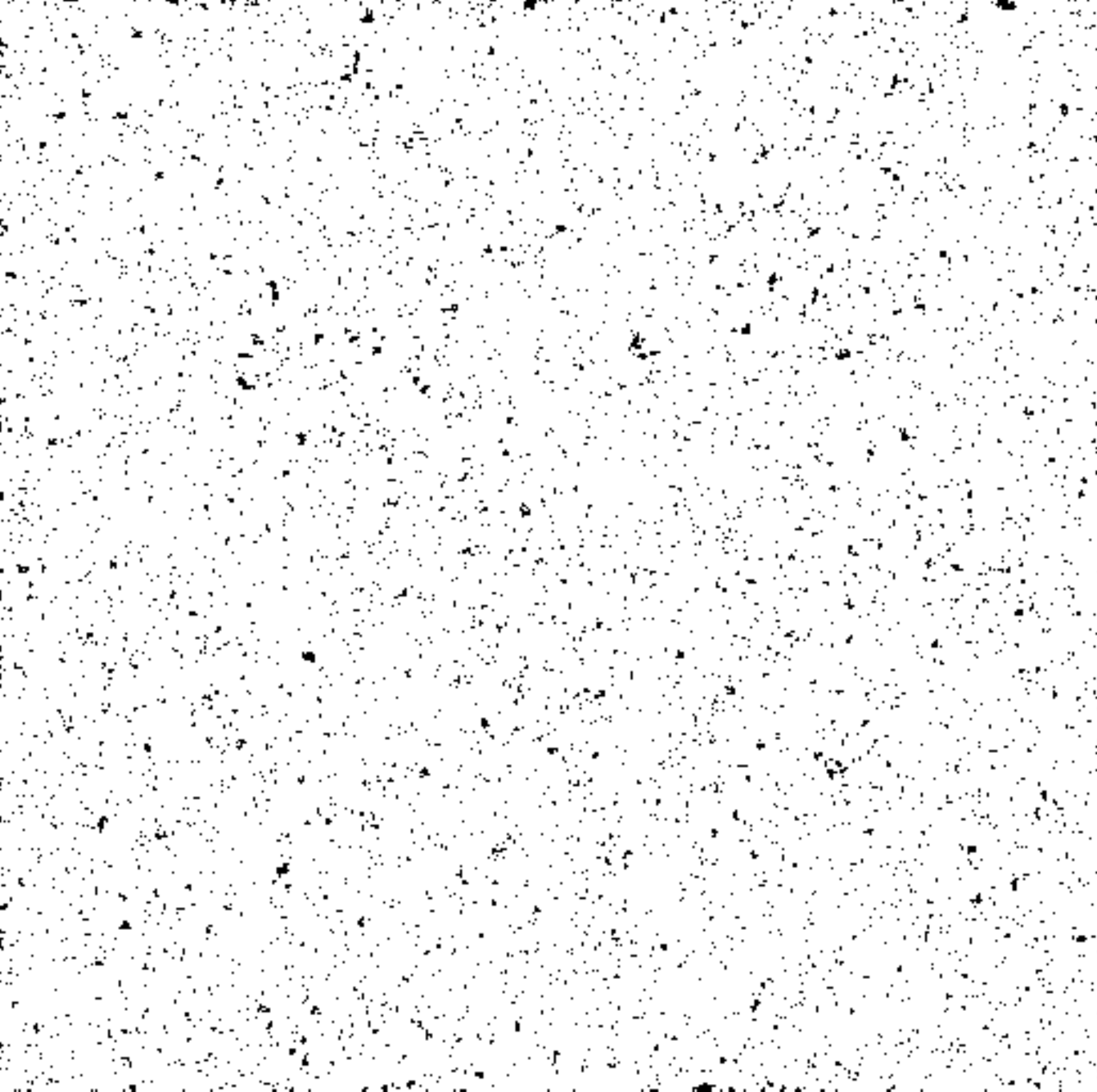}
\includegraphics[width=0.33\columnwidth,angle=90]{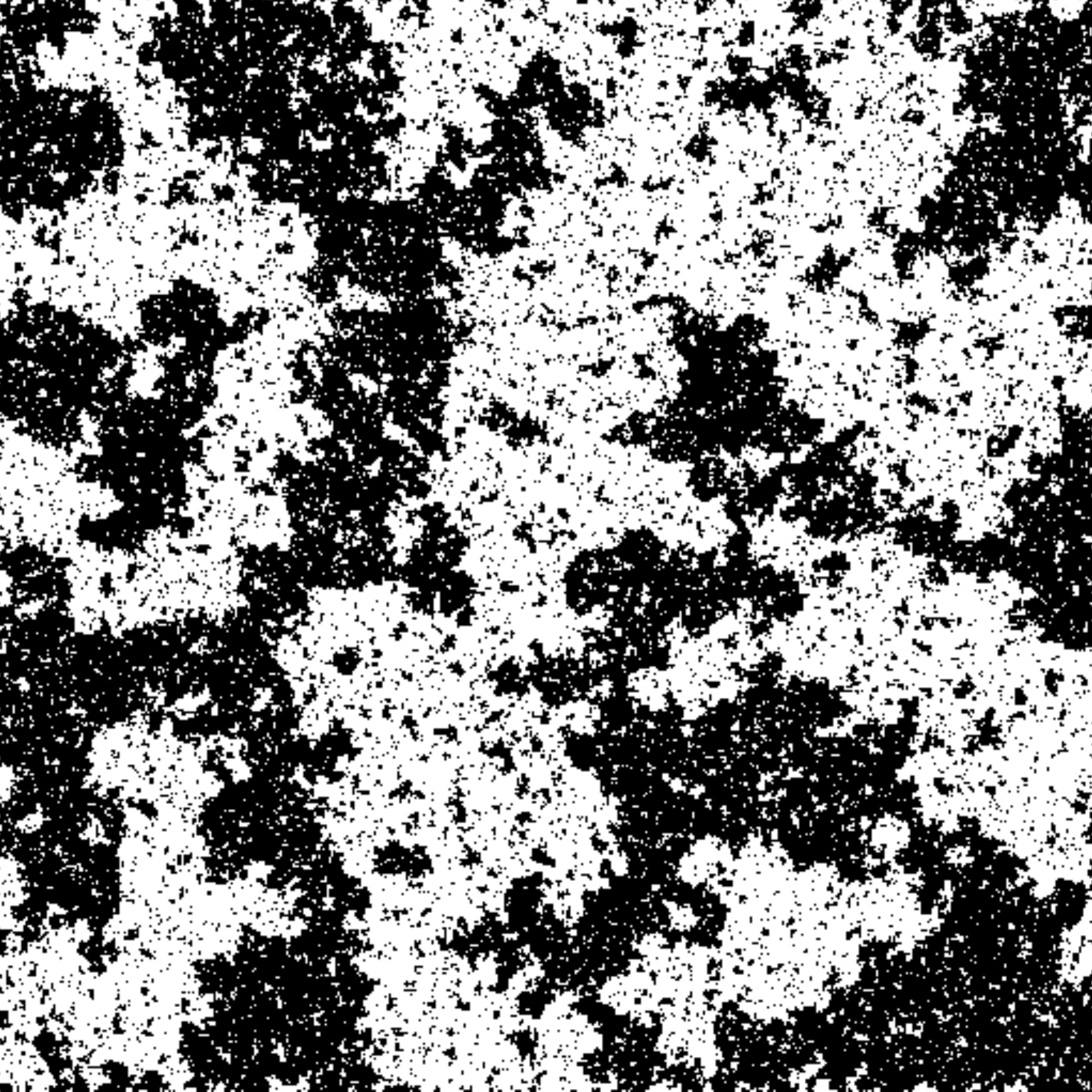}
\includegraphics[width=0.33\columnwidth,angle=90]{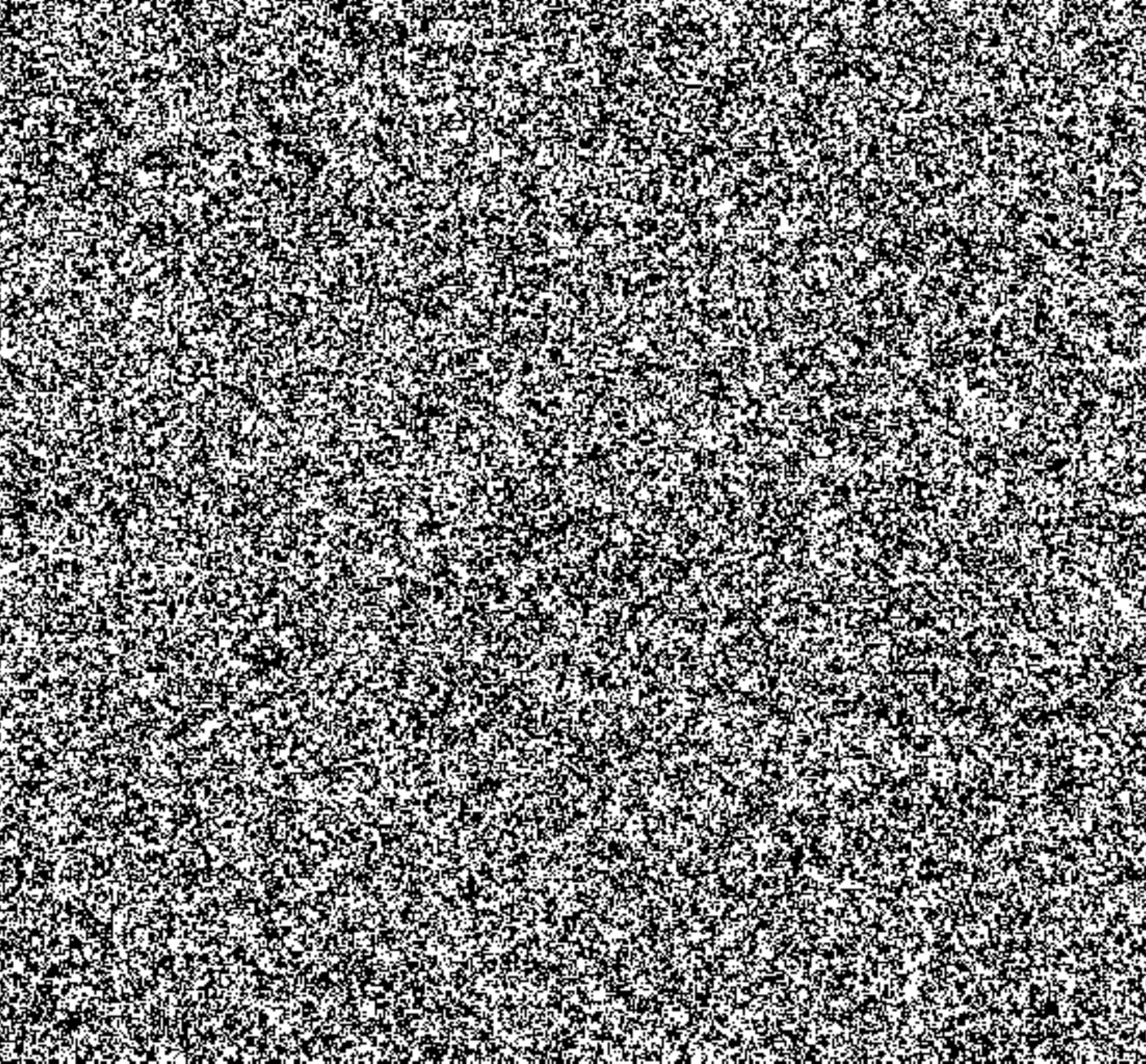}
\end{center}
\caption{Typical states of the Ising model on a $512 \times 512$ lattice at three different temperatures.  When $J/T$ is large enough (left) there are long-range correlations between the vertices; there are islands of the ``wrong'' state, but the fraction in the ``right'' state is bounded above $1/2$.  When $J/T$ is too small (right) the system is unmagnetized: correlations between vertices decay exponentially with distance, and in the limit $n \to \infty$ half of the vertices point in each direction.  
(From~\cite{moore-mertens-book}.)\label{fig:ising}}
\end{figure}

\section{Temperature and Noise} 

Pierre Curie discovered that iron has a phase transition where, at a critical temperature, its magnetization suddenly drops to zero because an equal fraction of atoms point up and down (see Figure~\ref{fig:ising} for the Ising model on the square lattice).  When the system is too hot and noisy, or equivalently if the interactions between neighbors are too weak, the correlations between atoms decay exponentially with distance, and the fraction of atoms pointed in each direction tends to $1/2$ as $n \to \infty$.  In the same way, in community detection we expect phase transitions where, if the community structure is too weak or the graph too sparse, the fraction of vertices labeled correctly will tend to $1/q$, no better than chance.

\pagebreak 
To make this precise, for each vertex $i$ define its \emph{marginal distribution}, i.e., the total probability that it belongs to a particular state $r$ according to the posterior distribution:
\begin{equation}
\label{eq:marginal}
\psi^i_r = P( \sigma_i = r \mid G ) = \sum_{\sigma: \sigma_i = r} P(\sigma \mid G) \, . 
\end{equation}
We can think of the magnetization as the average, over all vertices, of the probability that they are labeled correctly.  Rather than the ground state, another way to estimate $\sigma$ is to assign each vertex to its most likely group, 
\[
\sigma^*_i = \argmax_r \,\psi^i_r \, .
\]
This estimator, if we can compute it, maximizes the expected fraction of vertices labeled correctly~\cite{devroye-book,Iba99}.  
Thus the best possible algorithm for weak reconstruction is to find the marginals.  If they are uniform, so that every vertex is equally likely to belong to every group, the system is unmagnetized, and even weak reconstruction is impossible.  (The alert reader will object that, due to the permutation symmetry between the groups, the marginals $\psi^i$ are uniform anyway.  You're right, but we can break this symmetry by fixing the labels of a few vertices.)

\pagebreak
Let's pause here, and note that we are crossing a cultural divide.  The probability $P(G | \sigma)$ is a perfectly good objective function, and maximizing it gives the maximum likelihood estimator $\hat{\sigma}$.  This is the single most likely community structure the graph could have.  From a computer science point of view, solving this optimization problem seems like the right thing to do.  What could be wrong with that?

What we are saying is that rather than focusing on the ground state $\hat{\sigma}$, we need to think about the entire ``energy landscape'' of possible community structures.  If $\hat{\sigma}$ is only one of many local optima that have nothing in common with each other, it doesn't tell us anything about the ground truth---it is just overfitting, reacting to noise in the data rather than to an underlying pattern, like the bisections in Figure~\ref{fig:3reg}.  In a bumpy landscape with many hills, one hill is probably a little higher than the others.  If this hill corresponds to the best way to build a rocket, you should use it.  But if it corresponds to a hypothesis about a noisy data set, how confident should you be?

Bayesian inference demands that we understand the posterior distribution of our model, not just its maximum, and embrace the fact that we are uncertain about it.  The most likely single state in a ferromagnet is when all the atoms are aligned, but this is a bad guide to iron's physical properties.  Focusing on the ground state is like pretending the system is at absolute zero, which corresponds to assuming that $G$ is much more structured than it really is.  By focusing on the marginals instead, we ask what the likely $\sigma$ agree on.  
When data is sparse or uncertain, the average of many likely fits of a model is often better than the single ``best'' fit.

How can we compute these marginals, and thus the ``magnetization'' of the block model?  One approach, popular throughout machine learning and computational physics, is  Monte Carlo sampling.  We can flip the state of one vertex at a time, where the probability of each move depends on the ratio between the new and old values of the posterior~\eqref{eq:posterior}, or equivalently how much this move would raise or lower the energy.  After a long enough time, the resulting states are essentially samples from the equilibrium distribution, and taking many such samples gives us good estimates of the marginals.  This raises wonderful issues about how long it takes for this Markov chain to reach equilibrium, phase transitions between fast and slow mixing, and so on~\cite{LevinPeresWilmer2006}.  

Instead, we will use a method that tries to compute the marginals directly.  In addition to being an algorithm that is efficient and accurate under some circumstances, it lets us calculate analytically where phase transitions occur, and gives a hint of why, in some regimes, community detection might be possible but exponentially hard.

\section{The Cavity Method and Belief Propagation}

Each atom in a material is affected by its neighbors, and it affects them in turn.  In the Ising model, the probability that an atom points up depends on the probability that each of its neighbors does.  If we start with an initial estimate of these probabilities, we can repeatedly update each atom's probability based on those of its neighbors.  Hopefully, this process converges to a fixed point, which (also hopefully) gives the correct probabilities at equilibrium.  

This idea is called the \emph{cavity method} in physics~\cite{mezard-parisi-virasoro}, where it was invented to solve spin glasses---systems with random mixtures of ferromagnetic and antiferromagnetic interactions, such as the Sherrington-Kirkpatrick model~\cite{sherrington-kirkpatrick}.  But it turns out to be deeply related to message-passing algorithms such as belief propagation that were developed in the AI community a few years earlier, including in the Turing Award-winning work of Judea Pearl~\cite{pearl}.

As a warm-up, suppose you are a vertex $i$, and you know the marginals~\eqref{eq:marginal} of your neighbors. Using Bayes' rule as before, the fact that an edge exists between you and each of your neighbors affects your own marginal: for each neighbor $j$ in group $s$, the edge $(i,j)$ multiplies the relative probability that you are in group $r$ by $p_{rs}$.  Let's make the wild assumption that your neighbors' labels are independent, i.e., that their joint probability distribution is just the product of their marginals.  Then your own marginal is given by
\begin{equation}
\label{eq:naive}
\psi^i_r \propto \prod_{j: (i,j) \in E} \sum_{s=1}^q \psi^j_s \,p_{rs} \, , 
\end{equation}
where $\propto$ hides the normalization we need to make $\sum_{r=1}^q \psi^i_r = 1$.  We can iterate this equation, updating the marginal of a randomly chosen vertex at each step, until the marginals converge to a fixed point.\footnote{There could easily be $q!$ different fixed points, corresponding to permutations of the groups.  Any one of them will do.}

This approach is quite naive, and it is appropriately called ``naive Bayes.''  While it takes interactions into account in finding the marginals, once they are found it assumes the vertices are independent.  To put this differently, it assumes the posterior distribution is a product distribution, 
\[
P(\sigma \mid G) \approx \prod_{i=1}^n P(\sigma_i \mid G) = \prod_{i=1} \psi^i_{\sigma_i} \, . 
\]

\begin{figure}
\begin{center}
\includegraphics[width=1.4in]{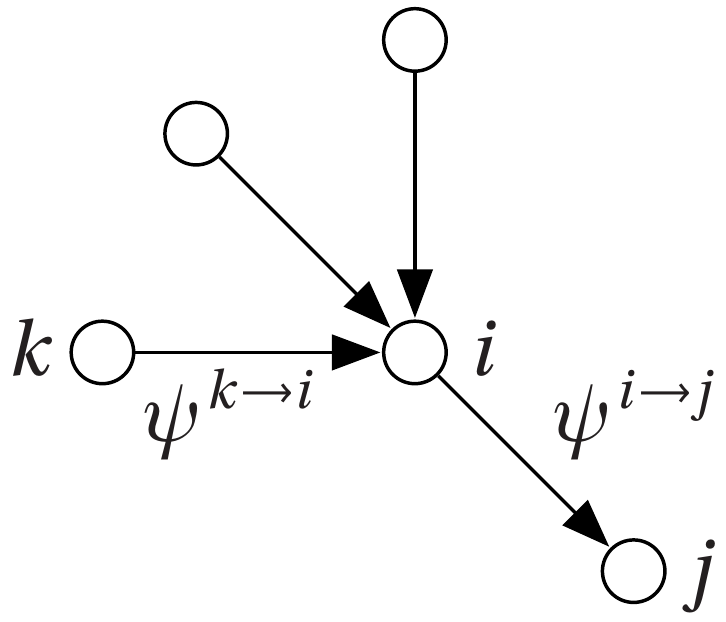}
\end{center}
\caption{In belief propagation, each vertex $i$ sends a message $\psi^{i \to j}$ to each of its neighbors $j$, consisting of an estimate of its marginal based on the messages it receives from its other neighbors $k \ne j$.\label{fig:bp}}
\end{figure}

Let's try something slightly less naive.  We will assume our neighbors are independent of each other, when conditioned on our own state.  Equivalently, we assume our neighbors are correlated only through us.  We can model this by having each vertex $i$ send each of its neighbors $j$ a ``message'' $\psi^{i \to j}$, which is an estimate of what $i$'s marginal would be if $j$ were not there---or more precisely, if we did not know whether or not there is an edge between $i$ and~$j$.  As shown in Figure~\ref{fig:bp}, $\psi^{i \to j}$ is an estimate of $i$'s marginal based only on $i$'s other neighbors $k \ne j$.  The update equation~\eqref{eq:naive} then becomes
\begin{equation}
\label{eq:bp}
\psi^{i \to j}_r \propto \prod_{\substack{k: (i,k) \in E \\ k \ne j}} \sum_{s=1}^q \psi^{k \to i}_s \,p_{rs} \, .
\end{equation}
The non-edges can be treated as a global interaction~\cite{Decelle2011} which we omit here.

Belief propagation consists of initializing the messages randomly and then repeatedly updating them with~\eqref{eq:bp}.  We typically do this asynchronously, choosing a vertex uniformly at random and updating its messages to all its neighbors.  If all goes well, this procedure converges quickly, and the resulting fixed point gives a good estimate of the marginals.  To compute the marginal of each vertex we use all its incoming messages, 
\begin{equation}
\label{eq:bp-marginal}
\psi^i_r \propto \prod_{j: (i,j) \in E} \sum_{s=1}^q \psi^{j \to i}_s \,p_{rs} \, .
\end{equation}
We define the two-point marginal $\psi^{ij}_{rs}$ as the joint distribution of $\sigma_i$ and $\sigma_j$, 
\[
\psi^{ij}_{rs} = P( \sigma_i=r \wedge \sigma_j=s \mid G ) \, ,
\]
and for neighboring pairs $i,j$ we estimate this from the messages as
\begin{equation}
\label{eq:bp-marginal2}
\psi^{ij}_{rs} \propto \psi^{i \to j}_r \psi^{j \to i}_s \,p_{rs} \, .  
\end{equation}

\pagebreak
Note that the message $i$ sends to $j$ does not depend on the message $j$ sends to $i$.  Why is this a good idea?  One intuitive reason is that it avoids an ``echo chamber'' where information bounces back and forth between two vertices, being pointlessly amplified.  It brings each vertex fresh information from elsewhere in the network, rather than just confirming its own beliefs.\footnote{At this point, I usually make a comment about talk radio and American politics.}  

Another reason to like~\eqref{eq:bp} is that it is exact on trees.  If the only paths between your neighbors go through you, then the conditional independence assumption is really true.  
If $G$ is not a tree, and your neighbors are connected by paths that don't go through you, we might still hope that conditional independence is a good approximation, if these paths are long and correlations decay rapidly with distance.  As a result, belief propagation can sometimes be shown to be asymptotically correct in ``locally treelike'' graphs, where most vertices do not lie on any short loops.  Standard counting arguments show that sparse graphs generated by the block model, like sparse \ER\ graphs, are in fact locally treelike with high probability.\footnote{People who deal with real networks know that they are far from locally treelike---they are full of short loops, e.g., because people introduce their friends to each other.  In this setting belief propagation is simply wrong, but in practice it is often not far off, and its speed makes it a useful algorithm in practice as long as the number of groups is small.}   

We can also derive~\eqref{eq:bp} by approximating the posterior with a distribution that takes correlations between neighbors into account, replacing the product distribution of each neighboring pair with its two-point marginal:
\begin{align}
\label{eq:bethe}
P(\sigma \mid G) 
&\approx \prod_{i=1}^n P(\sigma_i \mid G) 
 \prod_{(i,j) \in E} \frac{P(\sigma_i, \sigma_j \mid G)}{P(\sigma_i \mid G) P(\sigma_j \mid G)} \nonumber \\
&= \prod_{i=1}^n \psi^i_{\sigma_i} 
 \prod_{(i,j) \in E} \frac{\psi^{ij}_{\sigma_i,\sigma_j}}{\psi^i_{\sigma_i} \psi^j_{\sigma_j}} \, . 
\end{align}
Stable fixed points of~\eqref{eq:bp} are local minima, as a function of the messages, of the Kullback-Leibler divergence between the true posterior and~\eqref{eq:bethe}.  Equivalently, they are local minima of a quantity called the Bethe free energy~\cite{yedidia-freeman-weiss}.  But while~\eqref{eq:bethe} is exact if $G$ is a tree, for general graphs it doesn't even sum to one---in which case approximating the posterior this way is rather fishy.  

To my knowledge, belief propagation for the block model first appeared in~\cite{hastings}.  The authors of~\cite{Decelle2011,Decelle2011a} used it to derive, analytically but nonrigorously, the location of the detectability transition, and also to conjecture the detectable-but-hard regime where detection and reconstruction are information-theoretically possible but exponentially hard.\footnote{Some earlier physics papers also conjectured a detectability transition~\cite{reichardt-leone} and a hard regime~\cite{hu-ronhovde-nussinov} based on other methods, but did not locate these transitions precisely.}  They did this by considering the possible fixed points of belief propagation, whether they are stable or unstable, and how often random initial messages converge to them.  This brings us to the next section.

\section{Stability and Non-Backtracking Walks}

In the symmetric case~\eqref{eq:symmetric} of the block model, belief propagation has a trivial fixed point: namely, where all the messages are uniform, $\psi^{i \to j}_r = 1/q$.  If it gets stuck there, then belief propagation does no better than chance.  A good question, then, is whether this fixed point is stable or unstable.  If we perturb it slightly, will belief propagation fly away from it---hopefully toward the truth---or fall back in?  

The point at which this fixed point becomes unstable is called, among other things, the Kesten-Stigum threshold.  We will see that it occurs when 
\begin{equation}
\label{eq:kesten-stigum-cin-cout}
|\cin-\cout| > q \sqrt{c} \, , 
\end{equation}
where $c$ is the average degree~\eqref{eq:avgdeg}.  It was conjectured in~\cite{Decelle2011,Decelle2011a} that this is the computational threshold: that polynomial-time algorithms for weak reconstruction exist if and only if~\eqref{eq:kesten-stigum-cin-cout} holds.  Moreover, they conjectured that belief propagation is optimal, i.e., it achieves the highest possible accuracy.  

The scaling of~\eqref{eq:kesten-stigum-cin-cout} is intuitive, since it says that the expected difference in the number of neighbors you have inside and outside your group has to be at least proportional to the $O(\sqrt{c})$ fluctuations we see by chance.  What is interesting is that this behavior is sharp, and occurs at a specific constant.

The positive side of this conjecture is now largely proved.  For the case $q=2$, efficient algorithms for weak reconstruction when~\eqref{eq:kesten-stigum-cin-cout} holds were given in~\cite{massoulie2014,mns-proof}, and in~\cite{mns-colt} a form of belief propagation was shown to be optimal when $(\cin-\cout)/\sqrt{c}$ is sufficiently large.  For $q > 2$, another variant of belief propagation~\cite{abbe-sandon-more-groups} was shown to achieve weak reconstruction.  (Earlier papers~\cite{coja-mossel-vilenchik,coja-adaptive} had shown how to find planted colorings and bisections a constant factor above the threshold.)  Thus weak reconstruction is both possible and feasible above the Kesten-Stigum threshold.

In this section we describe how to locate the Kesten-Stigum threshold analytically.  Our discussion differs from the chronology in the literature, and has the clarity of hindsight; it will also introduce us to a class of spectral algorithms.  First suppose that the messages are almost uniform,
\[
\psi^{i \to j}_r = \frac{1}{q} + \eps^{i \to j}_r \, .
\]
If we substitute this in~\eqref{eq:bp} and expand to first order in $\eps$, the update equation becomes a linear operator on $\eps$, multiplying it by a matrix of derivatives:
\[
\eps := M \eps 
\quad \text{where} \quad
M_{((i,j),r),((k,\ell),s)} = \frac{\partial \psi^{i \to j}_r}{\partial \psi^{k \to \ell}_s} \, . 
\]
If $M$ has an eigenvalue whose absolute value is greater than $1$, the uniform fixed point is unstable.  If not, it is stable, at least locally.  

Determining $M$'s eigenvalues might seem a little daunting, since it is a $2mq$-dimensional matrix.  But it can be written quite compactly in terms of two simpler matrices~\cite{coja-mossel-vilenchik,Decelle2011,non-backtracking}.  First we consider the effect of a single edge $(k,i)$.  If we have no other information, then $i$'s outgoing message to $j$ is the one it receives from $k$, multiplied by a stochastic transition matrix $T$:
\[
\psi^{i \to j} = T \psi^{k \to i} 
\quad \text{where} \quad
T_{rs} = \frac{p_{rs}}{\sum_{r'} p_{r's}}
\]
This is again just Bayes' rule, and $T$ is just the matrix $p$ normalized so that its columns sum to $1$.  In the symmetric case~\eqref{eq:symmetric} this becomes
\begin{equation}
\label{eq:t-lambda}
T = \frac{1}{qc} \begin{pmatrix} 
\cin &\ldots &\cout \\ 
\vdots &\ddots \\
\cout & & \cin 
\end{pmatrix}
= \lambda \id + (1-\lambda) \frac{J}{q} \, , 
\end{equation}
where $\id$ is the $q$-dimensional identity matrix, $J$ is the $q \times q$ matrix of all $1$s, 
and 
\begin{equation}
\label{eq:lambda}
\lambda = \frac{\cin-\cout}{qc} 
\end{equation}
is the second eigenvalue of $T$.  Probabilistically, we can interpret~\eqref{eq:t-lambda} as copying $k$'s label to $i$ with probability $\lambda$, and choosing $i$'s label uniformly at random---just as if this edge did not exist---with probability $1-\lambda$.

Next, we define the following $2m$-dimensional matrix, whose rows and columns correspond to directed edges:
\[
B_{(i,j),(k,\ell)} = \begin{cases} 
1 & \mbox{if $\ell=i$ and $k \ne j$} \\
0 & \mbox{otherwise} \, . 
\end{cases} 
\]
We call this the non-backtracking matrix; in graph theory it is also called the Hashimoto matrix~\cite{hashimoto1989zeta}.  It corresponds to walks on $G$ where we are allowed to move in any direction except the one we just came from.  We can move from $k \to i$ to $i \to j$, but not flip the arrow to $i \to k$.  

A few lines of algebra show that
\[
M = B \otimes T \, . 
\]
In other words, $M$ is the $2mq$-dimensional matrix formed by replacing each $1$ in $B$ with a copy of $T$.  The appearance of $B$ encodes the non-backtracking nature of belief propagation, where $\psi^{i \to j}$ depends only on the incoming messages $\psi^{k \to i}$ for $k \ne j$.  

Now, the eigenvalues of the tensor product $B \otimes T$ are the products of eigenvalues of $B$ and $T$, so the question is whether any of these products exceed $1$.  However, there are a few eigenvalues that we can ignore.  First, being a stochastic matrix, $T$ has an eigenvalue $1$, with the uniform eigenvector $(1,\ldots,1)$.  A perturbation in this direction would mean increasing $\psi^{k \to i}_r$ for all $r$.  But since the messages are normalized so that $\sum_{r=1}^q \psi^{k \to i}_r = 1$, we must have $\sum_{r=1}^q \eps^{i \to j}_r = 0$.  Thus we can project the uniform vector away, leaving us just with $T$'s second eigenvalue $\lambda$.

What about $B$?  Given a Poisson degree distribution, an edge can move to $c$ new edges in expectation, so $B$'s leading eigenvalue is $c$.  However, the corresponding eigenvector $v$ is nonnegative and roughly uniform.  Perturbing the messages by $v \otimes w$ for a $q$-dimensional vector $w$ would correspond to changing all the marginals in the same direction $w$, but this would make some groups larger and others smaller.  At this point we (rather informally) invoke the non-edges that we've been ignoring up to now, and claim that they counteract this kind of perturbation---since if too many vertices were in the same group, the non-edges within that group would be very unlikely.  

It was conjectured in~\cite{non-backtracking}, and proved in~\cite{bordenave-lelarge-massoulie}, that with high probability $B$'s second eigenvalue approaches
\begin{equation}
\label{eq:mu}
\max(\mu,\sqrt{c}) 
\quad \text{where} \quad 
\mu = \frac{\cin-\cout}{q} = c\lambda \, .
\end{equation}  
To see this intuitively, consider a vector $v$ defined on directed edges $i \to j$ which is $+1$ if $i$ is in the first group, $-1$ if $i$ is in the second group, and $0$ if it is in any of the others.  Excluding the vertex it's talking to, each vertex has in expectation $\cin/q$ incoming messages from its own group and $\cout/q$ from each other group.  If these expectations held exactly, $v$ would be an eigenvector of $B$ with eigenvalue $\mu$.  The number of such messages fluctuates from vertex to vertex; but applying $B^\ell$ to $v$  smooths out these fluctuations, counting the total number of messages from each group entering a Poisson random tree of depth $\ell$ at its leaves.  Thus $B^\ell v$ converges to an eigenvector with eigenvalue $\mu$ as $\ell$ increases.

\begin{figure}
\begin{center}
\includegraphics[width=\columnwidth]{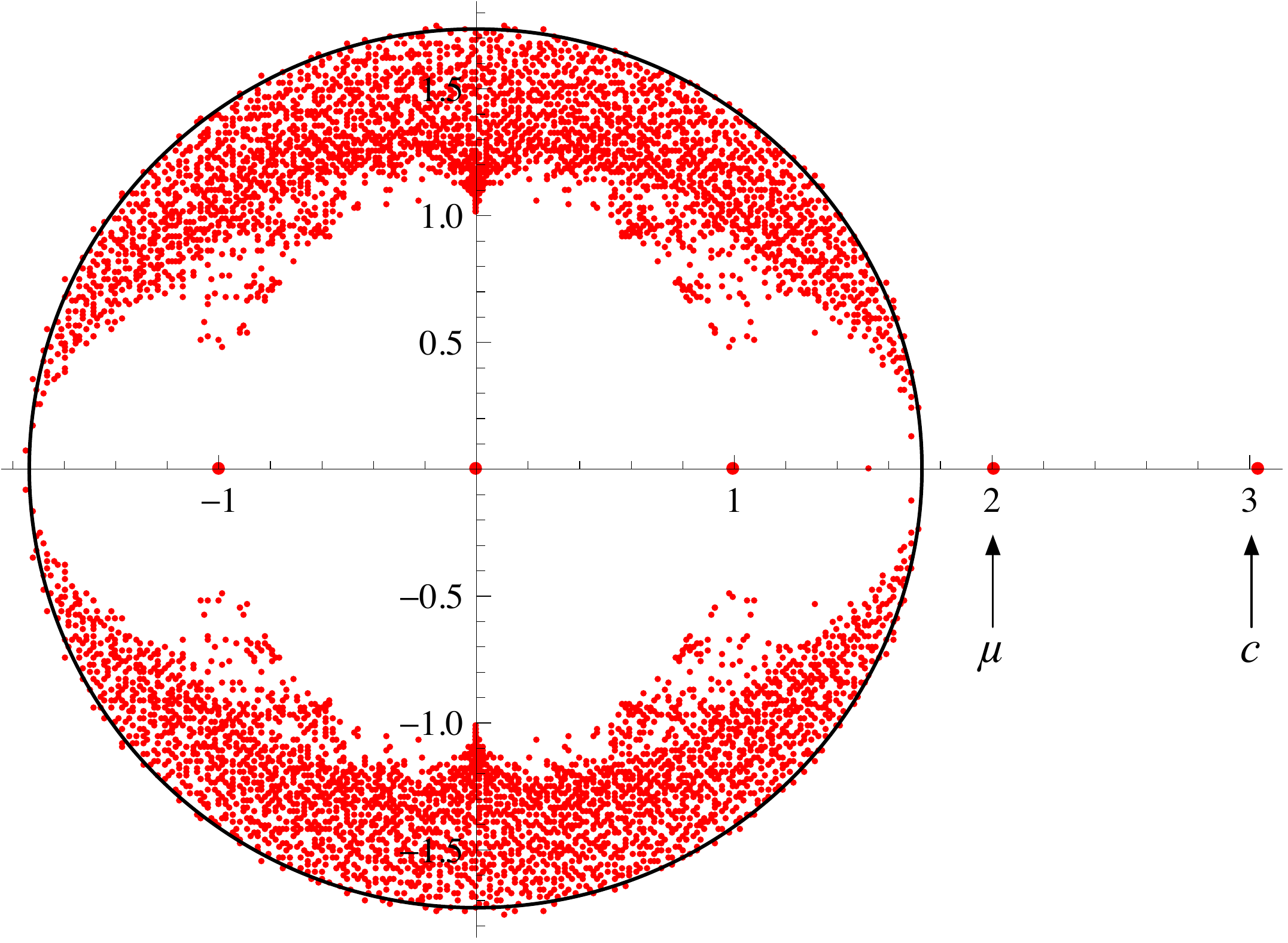}
\end{center}
\caption{The spectrum in the complex plane of the non-backtracking matrix of a graph generated by the stochastic block model with $n=4000$, $q=2$, $\cin=5$, $\cout=1$, and average degree $c=3$.  To the right on the real line we see the leading eigenvalues $c$ and $\mu=(\cin-\cout)/q=2$.  The eigenvector with eigenvalue $\mu$ is correlated with the communities whenever it is outside the bulk, so even ``linearized'' belief propagation---equivalently, spectral clustering with the non-bactracking matrix---will label nodes better than chance.\label{fig:spectrum}}
\end{figure}

Where does the $\sqrt{c}$ in~\eqref{eq:mu} come from?  You might enjoy showing that most eigenvalues have absolute value at most $\sqrt{c}$ by considering the trace of $B^\ell (B^T)^\ell$, which corresponds to taking $\ell$ steps forward and $\ell$ steps backward.  In fact, in Figure~\ref{fig:spectrum} you can see a ``bulk'' of eigenvalues inside the disk of radius $\sqrt{c}$ in the complex plane.  These eigenvectors come from the randomness in the graph, and are uncorrelated with the community structure.  

In any case, the relevant eigenvalue of $M = B \otimes T$ is thus $\mu \lambda$, and the uniform fixed point becomes unstable when
\begin{equation}
\label{eq:kesten-stigum-simultaneous}
\mu \lambda > 1 \quad \text{or} \quad \sqrt{c} \lambda > 1 \, ,
\end{equation}
both of which give 
\begin{equation}
\label{eq:kesten-stigum}
c \lambda^2 > 1 \, .
\end{equation}
This is the Kesten-Stigum threshold, and applying~\eqref{eq:lambda} gives~\eqref{eq:kesten-stigum-cin-cout}.  

If $\mu > \sqrt{c}$, which is also equivalent to~\eqref{eq:kesten-stigum}, then the 2nd through $q$th eigenvectors emerge from the bulk and become correlated with the communities, letting us label vertices according to their participation in these eigenvectors.  This suggests that we can skip belief propagation entirely, and use the non-backtracking matrix to perform spectral clustering.  

This spectral algorithm was proposed in~\cite{non-backtracking} and proved to work in~\cite{bordenave-lelarge-massoulie}.  While it is not quite as accurate as belief propagation, it performs better than chance.  (We note that the first algorithms to achieve weak reconstruction above the Kesten-Stigum threshold~\cite{massoulie2014,mns-proof} also worked by counting non-backtracking or self-avoiding walks.)  Roughly speaking, if we apply a small random perturbation to the uniform fixed point, the first few steps of belief propagation---powering the matrix of derivatives $M$---already point in a direction correlated with the planted assignment.  

In contrast, standard spectral clustering algorithms, using the adjacency matrix or graph Laplacian, fail in the constant-degree regime due to localized eigenvectors around high-degree vertices~\cite{non-backtracking}.  On the other hand, if the average degree grows moderately with $n$ so that the classic ``semicircle law'' of random matrix theory takes over, the Kesten-Stigum threshold can also be derived by calculating when a community-correlated eigenvector crosses the boundary of the semicircle~\cite{nadakuditi-newman}.


It may seem to the reader like a coincidence that two events happen simultaneously: $c \lambda^2 = 1$, so that the uniform fixed point becomes unstable, and $\mu = \sqrt{c}$, so that the second eigenvalue emerges from the bulk.  In fact, while $\mu$ depends only on the graph, $\lambda$ depends on the parameters we use to run belief propagation.  These thresholds coincide when we know the correct parameters of the block model that generated the graph.  If we run belief propagation with the wrong parameters, they become separate events, leading to a more complicated set of phase transitions.  For instance, if we assume that $\lambda$ is larger than it really is---assuming too much structure in the data---there appears to be a ``spin glass phase'' where the uniform eigenvector is unstable but the non-backtracking matrix tells us nothing about the ground truth, and belief propagation fails to converge~\cite{zhang-moore-pnas}.

Finally, another class of algorithms for weak reconstruction, which I will not discuss, use semidefinite relaxations.  These succeed almost all the way down to the Kesten-Stigum threshold~\cite{montanari-sen,javanmard-montanari-ricci-tersenghi}: specifically, they succeed at a threshold $c \lambda^2 = f(c)$ where $f(c)$ rapidly approaches $1$ as $c$ increases.

\section{Reconstruction on Trees}

In the previous section, we described efficient algorithms for weak reconstruction above the Kesten-Stigum threshold, proving the positive side of the conjectures made in~\cite{Decelle2011,Decelle2011a}.  What about the negative side?  For $q=2$, the same papers conjectured that, below the Kesten-Stigum threshold, even weak reconstruction is impossible, and no algorithm can label vertices better than chance.  This is a claim about information, not computation---that the block model is simply too noisy, viewed as a communication channel, to get information about the planted assignment from Nature to us.  In other words, for $q=2$ the computational and information-theoretic thresholds coincide.


This was proved in~\cite{mossel-neeman-sly-impossible} by connecting the block model to a lovely question that first appeared in genetics and phylogeny.  It is called reconstruction on trees, and it is where the Kesten-Stigum threshold was first defined~\cite{KestenStigum66}.  We will deal with it heuristically; see~\cite{mossel-peres-survey} for a survey of rigorous results.

\begin{figure}
\begin{center}
\includegraphics[width=\columnwidth]{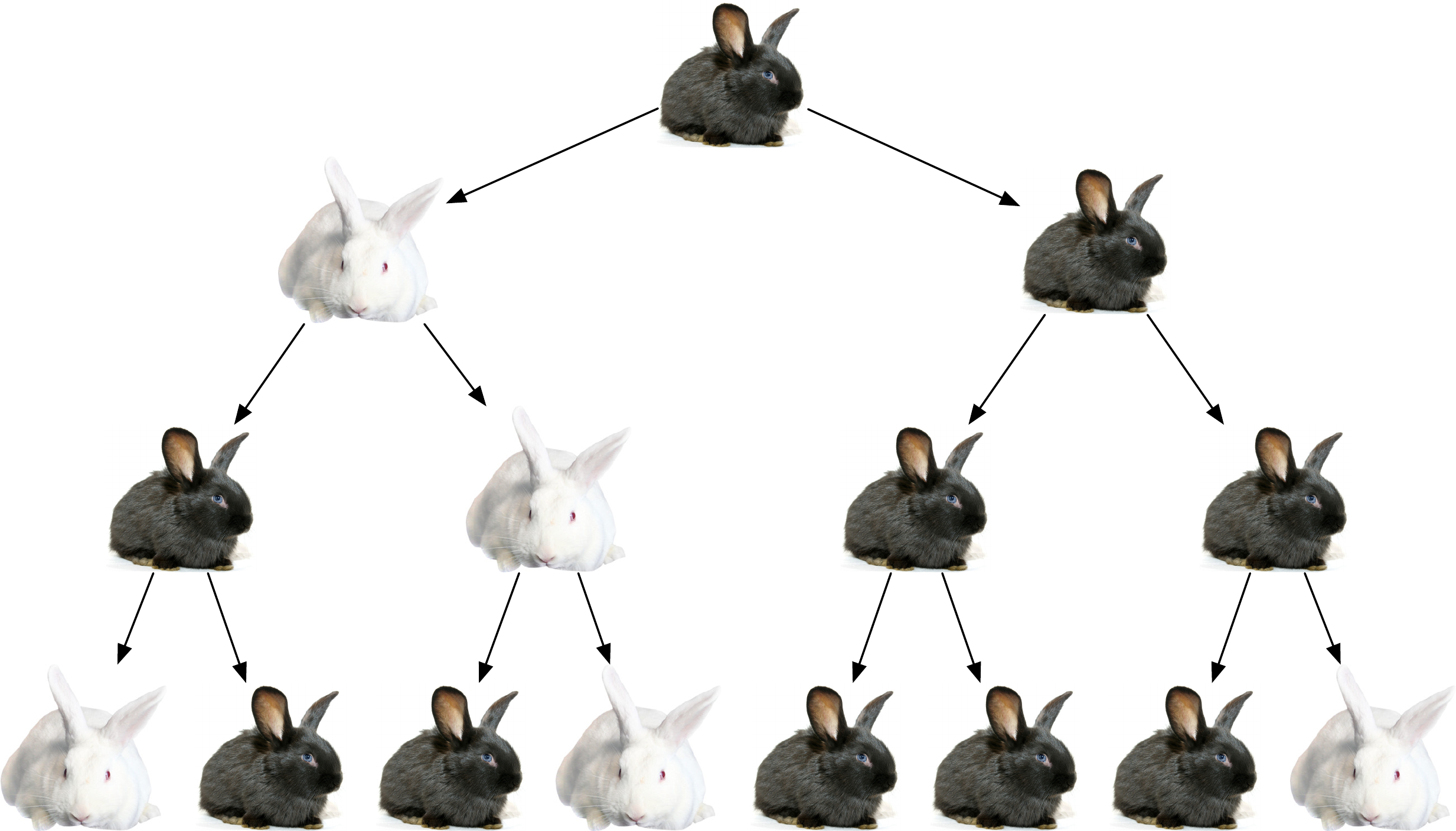}
\end{center}
\caption{The reconstruction problem on trees.  With probability $\lambda$, the color is copied from parent to child; with probability $1-\lambda$, the child's color is uniformly random.  Many generations later, the majority of the population has the same color as the root with high probability if and only if $c\lambda^2 > 1$.  The rabbits in this experiment have $c=2$ children at each step and $\lambda = 0.28$, below the Kesten-Stigum threshold.\label{fig:rabbits}}
\end{figure}

Consider a population of black and white rabbits.  They reproduce asexually~\cite{fibonacci}, and each one has $c$ children.  With probability $\lambda$, the parent's color is copied to the child; with probability $1-\lambda$, this information is lost and the child's color is uniformly random.  If we start with one rabbit, can we tell what color it was by looking at the population many generations later?  

After $\ell$ generations, there are $c^\ell$ rabbits.  The probability that a given one has an unbroken chain of inheritance, where the color was successfully passed from parent to child down through the generations, is $\lambda^\ell$.  Let us call such a descendant ``faithful.''  The expected number of faithful descendants is $c^\ell \lambda^\ell$.  If there are enough of them, then we can guess the progenitor's color by taking the majority of the population.

How many faithful descendants do we need?  One or two is not enough.  Through no fault of their own, each unfaithful descendant has a uniformly random color.  If we pretend they are independent, then the black ones will typically outnumber the white ones, or vice versa, by $\Theta(\sqrt{c^\ell})$.  The number of faithful descendants must be significantly larger than this imbalance for them to be in the majority with high probability.  Thus we need
\begin{equation}
\label{eq:ks-trees}
c^\ell \lambda^\ell \gg \sqrt{c^\ell} 
\quad \text{or} \quad
c \lambda^2 > 1 \, , 
\end{equation}
the same threshold we derived for the instability of the uniform fixed point. 
On the other hand, if $c \lambda^2 < 1$, then the number of faithful descendants is exponentially smaller than the typical imbalance, and the correlation between the majority and the progenitor's color decays exponentially in $\ell$.

Of course, it is not obvious that simply taking the majority is the best way to infer the progenitor's color.  The optimal strategy, by definition, is Bayesian inference, propagating the probabilities of the colors from the leaves to the root using---you guessed it---belief propagation.  But for $q=2$ the condition~\eqref{eq:ks-trees} is known to be both necessary and sufficient~\cite{evans-etal}.  It is also the precise threshold for \emph{robust} reconstruction, where the colors of the final population are given to us in a noisy way~\cite{janson-mossel}.

The analogy with the block model should be clear.  Above we showed~\eqref{eq:t-lambda} that each edge $(i,j)$ copies $i$'s label to $j$ with probability $\lambda$ and sets $j$'s label uniformly at random with probability $1-\lambda$.  Now imagine you want to infer the label of a vertex $i$.  Think of its neighbors as its children, their neighbors as grandchildren, and so on.  I generously give you the true labels of the vertices $\ell$ steps away---that is, its descendants after $\ell$ generations---and you try to infer the labels in the interior of this ball.  Below the tree reconstruction threshold, the amount of information you can gain about $i$ is exponentially small.  Thus the tree reconstruction threshold is a lower bound on the information-theoretic threshold for weak reconstruction, and for $q=2$ this matches the Kesten-Stigum threshold~\cite{mossel-neeman-sly-impossible}.

What happens for more than two groups?  For $q \ge 4$ the threshold is not known, but simply taking the plurality of the population is not optimal, so the Kesten-Stigum threshold is not tight~\cite{mossel-beating,mezard-montanari-reconstruction,sly-reconstruction}.  For $q=3$ it is believed to be tight, but as yet we have no proof of this.  As we will see, for $q \ge 4$ the information-theoretic and computational thresholds in the block model are distinct: in between, reconstruction is information-theoretically possible, but we believe it is computationally hard.    

\section{Detection and Contiguity}

We turn now from reconstruction to detection.  If $G$ is too sparse or the community structure is too weak, we can't even tell whether $G$ was generated by the block model or the \ER\ model.  What do we mean by this, and how can we prove it?

Let $P(G)$ and $Q(G)$ be the probability of $G$ in the block model and the \ER\ model $G(n,c/n)$ respectively.  These two distributions are not statistically close: that is, $|P-Q|_1 = \sum_G |P(G)-Q(G)|$ is not small.  If they were, they would be hard to tell apart even if we had many independent samples of $G$.  But we are only given one graph $G$, so we mean that $P$ and $Q$ are close in a weaker sense.

Imagine the following game. I flip a coin, and generate $G$ either from the block model or from the \ER\ model.  I then show you $G$, and challenge you to tell me which model I used.  It is easy to see that you can do this with probability greater than $1/2$, simply by counting short loops.  For instance, in the limit $n \to \infty$ the expected number of triangles in $G(n,c/n)$ is $c^3/6$, while in the block model it is (exercise!) 
\[
\frac{1}{6q^3} \,n^3 \,\tr\ p^3 = \frac{c^3}{6} \left( 1 + (q-1)\lambda^3 \right) \, , 
\]
where $p$ is the $q \times q$ matrix in~\eqref{eq:sbm-general} and $\lambda$ is defined as in~\eqref{eq:lambda}.  In both models the number of triangles is asymptotically Poisson, so these distributions overlap, but they are noticeably different for any $\lambda > 0$.  A similar calculation~\cite{mossel-neeman-sly-impossible} shows that above the Kesten-Stigum threshold, the number of cycles of length $\ell$ has less and less overlap as $\ell$ increases.

When we say that detection is impossible, we mean that you can't win this game with probability $1-o(1)$.  To prove this, we need a concept called \emph{contiguity}.  Let $P$ and $Q$ be two distributions, or to be more precise, families of distributions $(P_n), (Q_n)$ on graphs of size $n$.  We write $P \contig Q$ if, for any event $E$ such that $\lim_{n \to \infty} Q(E) = 0$, we also have $\lim_{n \to \infty} P(E) = 0$.  Turning this around, any event $\overline{E}$ that holds with high probability in $Q$ also holds with high probability in $P$.  We say that $P$ and $Q$ are mutually contiguous if $P \contig Q$ and $Q \contig P$.  

This idea first appeared in random graph theory, in the proof that random 3-regular graphs have Hamiltonian cycles with high probability.  One way to make 3-regular graphs is to start with a random $n$-cycle, add a random perfect matching, and condition on having no multiple edges.  These graphs are Hamiltonian by construction---indeed, this is precisely a planted model---and this distribution can be shown to be contiguous to the uniform distribution on 3-regular graphs~\cite{robinson-wormald}.

If we can show that the block model $P$ and the \ER\ model $Q$ are contiguous, then any algorithm that correctly says ``yes, there are communities'' with high probability in the block model would have to (incorrectly) say that in the \ER\ model as well.  Therefore there is no algorithm which is correct with high probability in both cases.  

\pagebreak
This has some interesting scientific consequences.  Consider the Ising model, or a model of an epidemic, or any other dynamical process defined in terms of a graph.  If $P$ and $Q$ are contiguous, then any quantity produced by this process---the magnetization, the fraction of the population that becomes infected, whatever---must have overlapping distributions when $G$ is generated by $P$ or by $Q$.  In particular, if this quantity is tightly concentrated, then it must have the same typical value for both $P$ and $Q$.

How can we prove contiguity?  Here we will describe a sufficient condition that implies $P \contig Q$.  This is already enough to prevent the existence of an algorithm that distinguishes them with high probability, and thus prove a lower bound on the detectability transition.

Our starting point is inspired by statistics.  If you have a fancy model $P$ and a null model $Q$, a classic way to tell which one generated the data $G$---and therefore whether the structure $P$ describes is really there---is to compute the likelihood ratio $P(G)/Q(G)$ and reject the null hypothesis if it exceeds a certain threshold.  The Neyman-Pearson lemma~\cite{neyman-pearson} shows that this is the best possible test in terms of its statistical power.  

If $P$ and $Q$ were nearly disjoint, then $P/Q$ would be either very small or very large, letting us classify $G$ with high probability.  Conversely, if $P/Q$ is typically bounded, then $P$ and $Q$ are contiguous.  In fact, it's sufficient to bound the expectation of $P/Q$ when $G$ is drawn from $P$, or equivalently its second moment when $G$ is drawn from $Q$:
\begin{equation}
\label{eq:ratio-moments}
\Exp_P \!\left[ \frac{P}{Q} \right]
= \sum_G P(G) \,\frac{P(G)}{Q(G)} 
= \sum_G Q(G) \,\frac{P(G)^2}{Q(G)^2} 
= \Exp_Q \!\left[ \frac{P^2}{Q^2} \right] \, .
\end{equation}
Let $E$ denote the event that $G$ belongs to some set of graphs, and let $1_E$ denote the indicator random variable for this event.  If this second moment is bounded by a constant $C$, the Cauchy-Schwarz inequality gives 
\begin{align}
P(E) = \Exp_P [1_E] = \Exp_Q \!\left[\frac{P}{Q} \,1_E\right] 
\le \sqrt{ \Exp_Q \!\left[ \frac{P^2}{Q^2} \right] \,\Exp_Q [1_E^2] } 
\le \sqrt{ C Q(E) } \, .
\end{align}
Then $Q(E) \to 0$ implies $P(E) \to 0$, and a bounded second moment is enough to imply $P \contig Q$.  Proving $Q \contig P$ takes significantly more work; it uses a refinement of the second moment method pioneered in~\cite{robinson-wormald} that conditions on the number of short cycles.  We will not discuss it here.

\pagebreak
If the reader is curious, we can describe how to bound the second moment.  Since $P(G)$ is a sum over all possible group assignments~\eqref{eq:z}, expanding the second moment gives a sum over pairs of assignments,
\begin{align*}
\frac{P(G)^2}{Q(G)^2} 
&= \frac{1}{q^{2n}} \left( \sum_\sigma \frac{P(G \mid \sigma)}{Q(G)} \right)^2 
= \frac{1}{q^{2n}} \sum_{\sigma,\tau} \frac{P(G \mid \sigma) P(G \mid \tau)}{Q(G)^2} \\ 
&= \frac{1}{q^{2n}} \sum_{\sigma,\tau} \prod_{(i,j)} \begin{cases}
\displaystyle{\frac{p_{\sigma_i,\sigma_j} p_{\tau_i,\tau_j}}{p^2}} & \mbox{if $(i,j) \in E$} \\
\displaystyle{\frac{(1-p_{\sigma_i,\sigma_j})(1-p_{\tau_i,\tau_j})}{(1-p)^2}} & \mbox{if $(i,j) \notin E$} \, .
\end{cases} 
\end{align*}
By linearity of expectation, we can move the expectation through this sum.  Moreover, since each edge $(i,j)$ exists with probability $p$ in the \ER\ model, and these events are independent, we can move the expectation into the product.  Taking the sparse case $p_{rs} = c_{rs}/n$ and using the approximatinos $1+x \approx \e^x$ and $1/(1-x) \approx 1+x$, a few lines of algebra give
\begin{align}
\label{eq:second-p}
\Exp_Q \!\left[ \frac{P^2}{Q^2} \right] 
&= \frac{1}{q^{2n}} \sum_{\sigma,\tau} 
\prod_{(i,j)} \left( \frac{p_{\sigma_i,\sigma_j} p_{\tau_i,\tau_j}}{p} + \frac{(1-p_{\sigma_i,\sigma_j})(1-p_{\tau_i,\tau_j})}{1-p} \right) 
\nonumber \\
&\approx \frac{1}{q^{2n}} \sum_{\sigma,\tau} 
\exp\!\left[ \frac{c}{n} 
\sum_{(i,j)} \left( \frac{c_{\sigma_i,\sigma_j}}{c} - 1 \right) \left( \frac{c_{\tau_i,\tau_j}}{c} - 1 \right) \right] 
\, ,
\end{align}
where the $\approx$ hides a $\Theta(1)$ multiplicative error.

Now, the summand in~\eqref{eq:second-p} is a function only of the number of vertices assigned to each pair of groups by $\sigma$ and $\tau$.  Define a $q \times q$ matrix $\alpha$, where $\alpha_{rs}$ is $q$ times the fraction of vertices $i$ such that $\sigma_i=r$ and $\tau_i=s$.  If there are $n/q + o(n)$ vertices in each group, $\alpha$ is doubly stochastic.  Then in the symmetric case, a few more lines of algebra give
\[
\Exp_Q \!\left[ \frac{P^2}{Q^2} \right] 
= \frac{1}{q^{2n}} \sum_{\sigma,\tau} \exp\!\left[ \frac{c \lambda^2 n}{2} \left( |\alpha|_F^2 - 1 \right) \right] \, , 
\]
where $|\alpha|_F^2 = \tr (\alpha^T \alpha) = \sum_{rs} \alpha_{rs}^2$ denotes the Frobenius norm.

If $\sigma$ and $\tau$ are uncorrelated, then $\alpha_{rs} = 1/q$ for all $r,s$ and $|\alpha|_F^2 = 1$.  In that case, the summand is $1$.  But when $\sigma$ and $\tau$ are identical, $\alpha=\id$, $|\alpha|_F^2 = q$, and the summand is exponentially large.  The question is thus whether the uncorrelated pairs dominate the sum.  In physical terms there is a tug of war between entropy---the fact that most pairs of assigments $\sigma, \tau$ are uncorrelated---and energy, which favors correlated pairs.  

\pagebreak
We can approximate this sum using Laplace's method (e.g.~\cite{achlioptas-moore-ksat}).  For $q=2$ this gives a one-dimensional optimization problem; for larger $q$, it requires us to maximize a function over the space of doubly-stochastic matrices~\cite{achlioptas-naor}.  Happily, we find that when $\lambda$ is small enough, entropy wins out, and the second moment is bounded by a constant.

These sorts of calculations, and their refinement conditioned on the number of small cycles, were used in~\cite{mossel-neeman-sly-impossible} to prove contiguity below the Kesten-Stigum threshold for $q=2$, and in~\cite{banks-etal-colt} to prove contiguity as long as
\begin{equation}
\label{eq:banks}
c \lambda^2 < \frac{2 \ln (q-1)}{q-1} \, , 
\end{equation}
giving a lower bound on the information-theoretic detectability threshold for general $q$.  Separate arguments in those papers show that reconstruction is also information-theoretically impossible in this range.

To prove an upper bound on the information-theoretic threshold, it suffices to find a property---even one which takes exponential time to verify---that holds with high probability in the block model but not in $G(n,c/n)$.  One such property is the existence of a partition which is as ``good'' as the planted one in terms of the number of edges between groups.  A simple counting argument~\cite{banks-etal-colt} shows that when $c$ is sufficiently large, with high probability $G(n,c/n)$ has no such partition, and all such partitions in the block model are correlated with the planted one.  Thus we can achieve both detection and reconstruction with exhaustive search.  

Combined with~\eqref{eq:banks}, this determines the information-theoretic threshold to within a multiplicative constant.  It also shows that there are $\lambda$ for which the information-theoretic threshold is below the Kesten-Stigum threshold for $q \ge 5$.  In the disassortative case we knew this already; in the planted coloring model we have $\cin=0$ and $\lambda = -1/(q-1)$, so the Kesten-Stigum threshold is at $c = (q-1)^2$.  But a first-moment argument shows that the threshold for $q$-colorability in $G(n,c/n)$ is below this point for $q \ge 5$, so we can distinguish between the two models by searching exhaustively for a $q$-coloring.

A tighter argument that focuses on typical partitions rather than a single good one~\cite{abbe-sandon-isit-2016,abbe-sandon-more-groups-arxiv} shows that there exist $\lambda < 0$ such that the information-theoretic threshold is below the Kesten-Stigum threshold for $q \ge 4$.  This proves the conjecture of~\cite{Decelle2011} that these thresholds become distinct for $q \ge 4$ in the disassortative case.  The case $q=3$, where they are conjectured to coincide, remains open.

\pagebreak

\section{The Hard Regime}

Now that we know the Kesten-Stigum threshold and the information-theoretic thresholds are different for large enough $q$, what happens in between them?  And why do we believe that detection and reconstruction are exponentially hard in this regime?  Here's the physics picture---not all of it is rigorous yet, but rapid progress is being made.

\begin{figure}
\begin{center}
\includegraphics[width=0.8\columnwidth]{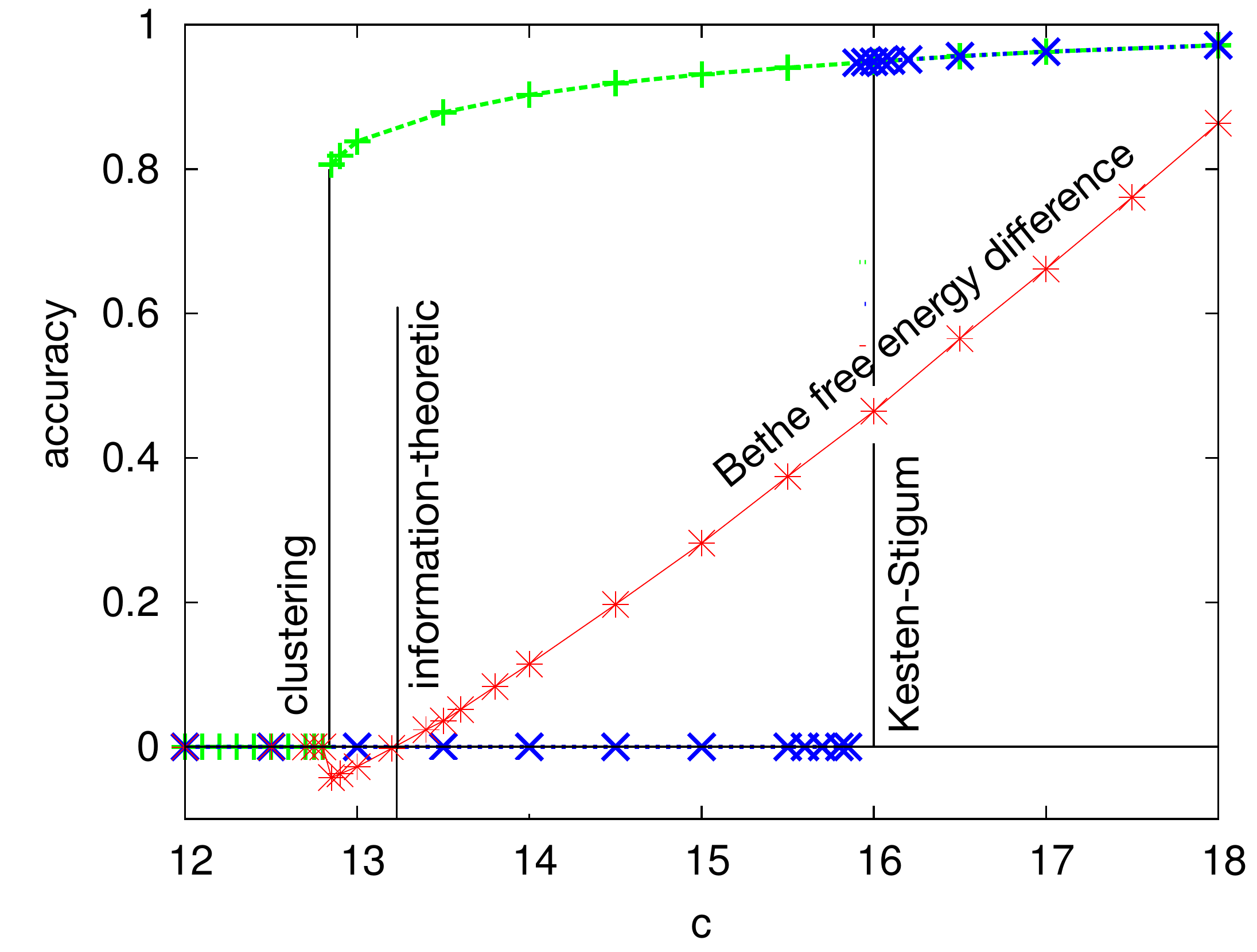}
\end{center}
\caption{The accuracy achieved by belief propagation in the planted coloring model with $q=5$ as a function of the average degree $c$, normalized so that $0$ is chance.  The blue curve shows the result of initializing the messages randomly; as expected, below the Kesten-Stigum threshold at $c=(q-1)^2=16$, the uniform fixed point is stable and most initial messages fall into it, giving an accuracy of zero (i.e., no better than chance).  The green curve shows another fixed point; it is quite accurate, but it has a very narrow basin of attraction.  The red curve shows the Bethe free energy difference between these two fixed points.  The information-theoretic detectability transition occurs at $c=13.2$ where this curve crosses zero.  The clustering transition, where tempting nonuniform fixed points first appear, takes place at $c=12.8$.  (From~\cite{Decelle2011}.)\label{fig:hard}}
\end{figure}

Take a look at Figure~\ref{fig:hard}, which is copied from~\cite{Decelle2011}.  It shows the result of belief propagation on the planted coloring model with $q=5$ groups.  (In this case the hard regime is particularly easy to see numerically.)  There are two curves, corresponding to two different fixed points: the uniform one, and an accurate one which is correlated with the planted assignment.  

Above the Kesten-Stigum threshold at $c=16$, the uniform fixed point is unstable.  Belief propagation flies away from it, and arrives at the accurate fixed point.  When $c < 12.8$, on the other hand, the accurate fixed point disappears and belief propagation always ends up at the uniform point.

In between these two values of $c$ the story is more interesting.  The uniform fixed point is locally stable, and it attracts most initial states.  If we initialize the messages randomly, we fall into it and perform no better than chance (the blue curve in Figure~\ref{fig:hard}).  But if we are lucky enough to start close to the accurate fixed point, we find it is locally stable as well, achieving the accuracy shown in green.  The problem is that the accurate fixed point has a very small basin of attraction---it only attracts an exponentially small fraction of initial messages.  Indeed, we were only able to find it numerically by cheating and initializing the messages with the ground truth.

If we had the luxury of exponential time, we could explore the entire space, and find both the accurate fixed point and the uniform one.  How would we choose between them?  In Bayesian terms, we want to know which one contributes the most to the total probability $Z=P(G)$ of the graph.  As we alluded to above, each fixed point $\psi$ of belief propagation has a quantity called the Bethe free energy: this is an estimate of $-(1/n) \log Z_\psi$, where $Z_\psi$ is the contribution to $Z$ from $\sigma$ distributed according to~\eqref{eq:bethe}.  If we believe this estimate, we will select the fixed point with the lowest Bethe free energy, since the corresponding $\sigma$ have exponentially more probability in the posterior distribution.  As shown in red in Figure~\ref{fig:hard}, the free energies of the uniform and accurate fixed points cross at $c = 13.2$.  This is the information-theoretic threshold: the point at which Bayesian inference, given full access to the posterior distribution, would find an accurate partition.  

\pagebreak
While detection is information-theoretically possible above this point, we believe that below the Kesten-Stigum threshold the accurate fixed point is exponentially hard to find.  Physically, it lies behind a ``free energy barrier,'' a bottleneck of high-energy, low-probability states that separate it from the uniform, unmagnetized state.  This is like a material whose lowest-energy state is a perfect crystal, but which almost always gets stuck in a glassy, amorphous configuration instead.  It would rather be in the crystalline state, but it would take longer than the age of the universe to find it.  Similar phenomena occur in low-density parity check codes~\cite{mezard-montanari-book}: in some regimes the correct codeword is hidden behind a free energy barrier, and message-passing algorithms for error correction get stuck in a glassy state.  

Of course, by ``hard'' here we do not mean that this problem is NP-hard.  It seems unlikely that a large class of other problems could be reduced to detection or reconstruction in the block model, even in an average-case sense.  But we could hope to prove that these problems are hard for specific classes of algorithms: for instance, that Markov Chain Monte Carlo algorithms take exponential time to find an assignment correlated with the planted one, or that (as seems to be the case) belief propagation converges to the uniform fixed point from all but an exponentially small fraction of initial messages.

Between $c=12.8$ and $c=13.2$ is a curious region.  The accurate fixed point still exists, but it has higher Bethe free energy than the uniform one.  Even if we knew about it, we would choose the uniform fixed point instead, and conclude that there aren't really any communities.  The accurate fixed point corresponds to a set or ``cluster'' of good-looking partitions, which agree with each other on most vertices; but there are exponentially many such clusters, separated from each other by a large Hamming distance, and like the bisections in Figure~\ref{fig:3reg} they would exist in a random graph as well.   Focusing on any one of them would put us in danger of overfitting.

These transitions have deep analogies with phenomena in random constraint satisfaction problems~\cite{mezard-montanari-book,krzakala-etal-gibbs,moore-mertens-book}.  The information-theoretic threshold is also the \emph{condensation} transition, above which a single cluster dominates the set of solutions; in a planted model, this cluster contains the planted state, and corresponds to the accurate fixed point.  The point at which the accurate fixed point first appears is the \emph{clustering} transition, also known as the \emph{dynamical replica symmetry breaking transition}, where the landscape of solutions becomes bumpy.  It is also the threshold for \emph{Gibbs extremality}, which is related to spatial decay of correlations in typical states, and for reconstruction on trees~\cite{mezard-montanari-reconstruction}.

Very recent work~\cite{coja-etal-cavity} shows that the Bethe free energy is asymptotically correct in some planted models, including the disassortative case of the block model, and rigorously identifies the information-theoretic threshold with the condensation transition~\cite{coja-new}.  This makes it possible, in principle, to locate the information-theoretic threshold exactly.  While the resulting calculation is not fully explicit, this gives a rigorous justification for numerical methods used in physics based on the cavity method.  See also~\cite{ding-sly-sun,sly-sun-zhang} for recent results on random satisfiability that justify the cavity method even beyond the condensation transition.  

\begin{figure}
\begin{center}
\includegraphics[width=0.9\columnwidth]{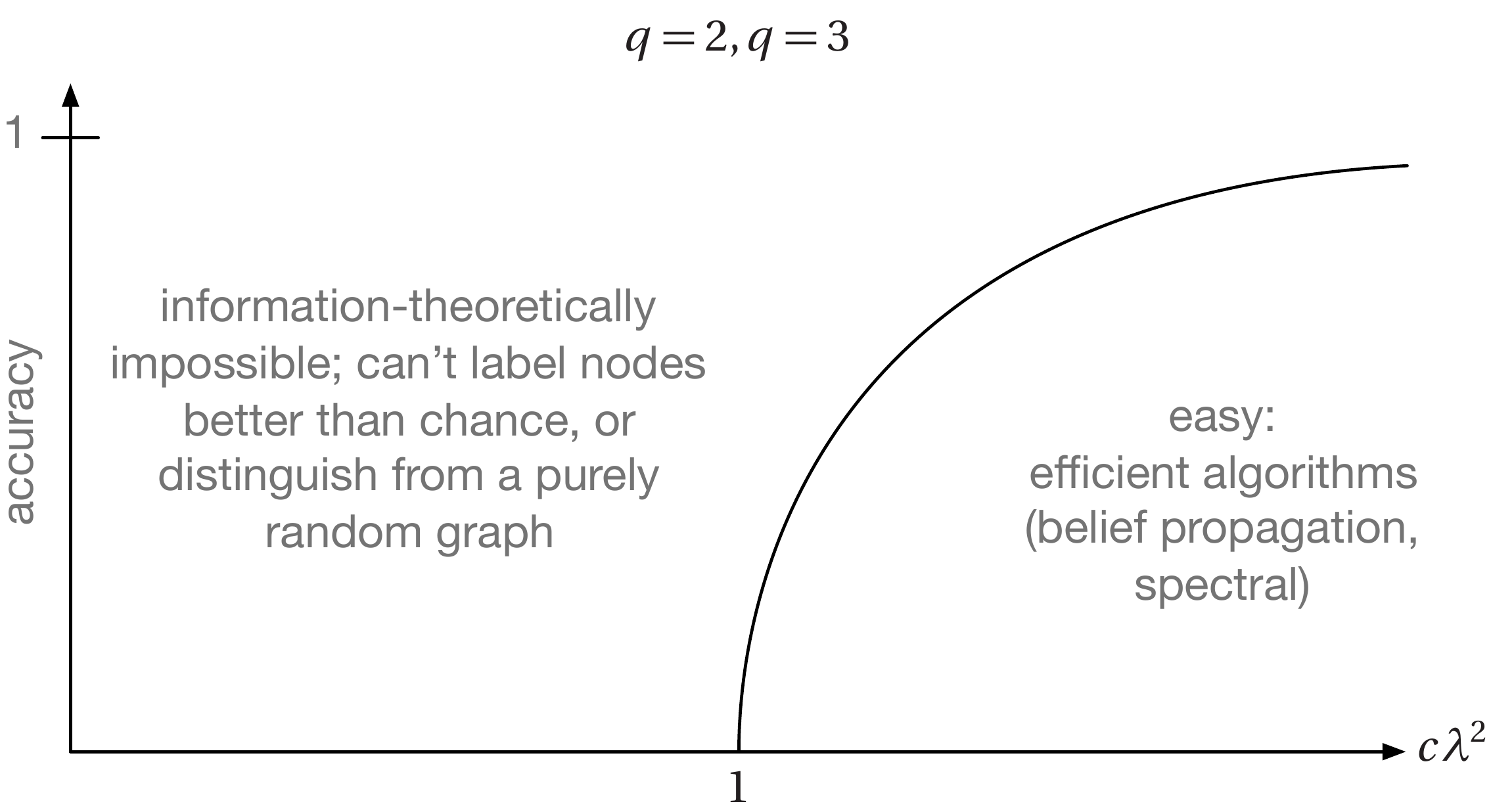} \\
\bigskip
\includegraphics[width=0.9\columnwidth]{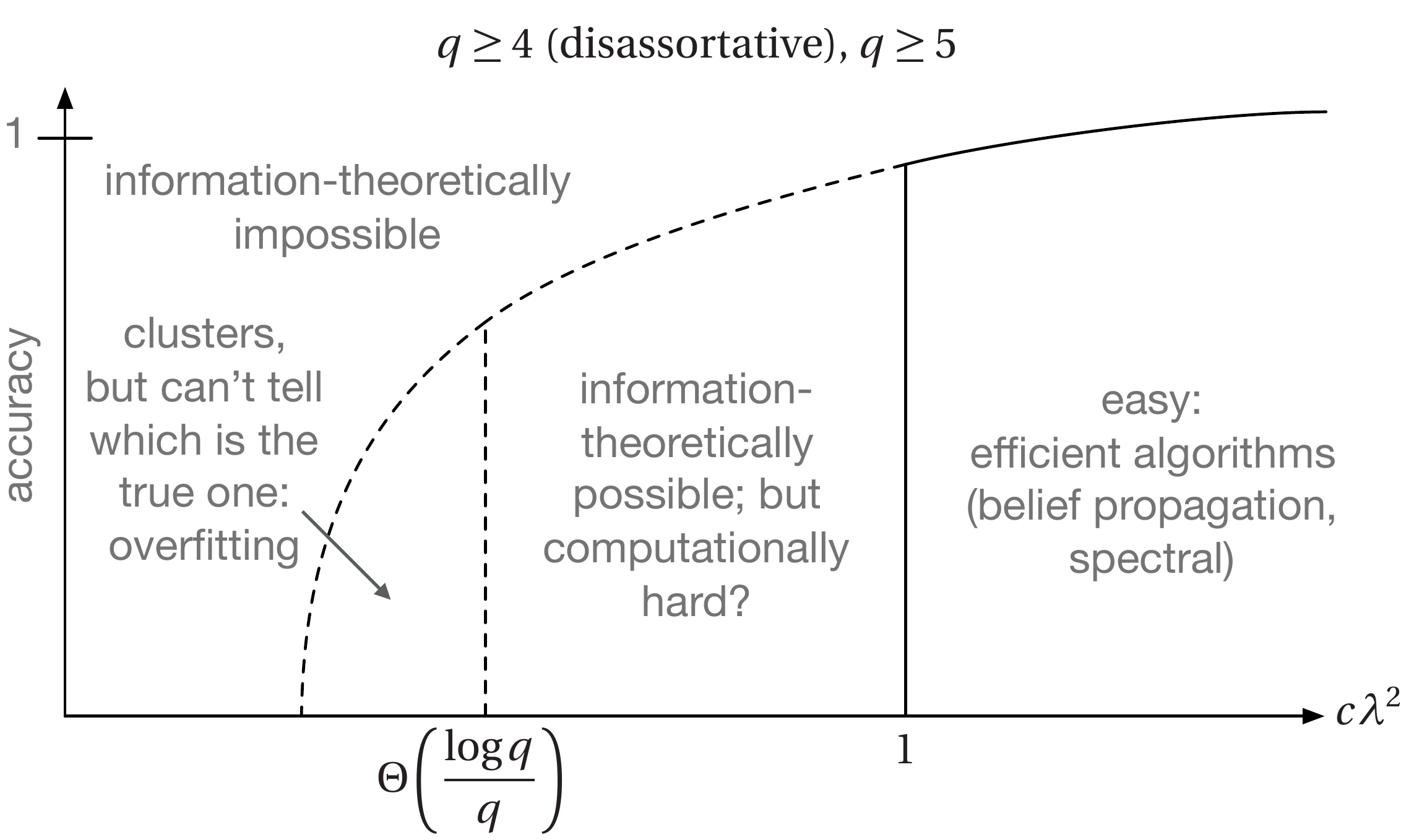} \\
\end{center}
\caption{A summary of known and conjectured phase transitions in the block model where the number of groups is $q=2, 3$ (top) and $q \ge 4$ (bottom).\label{fig:summary}}
\end{figure}

We summarize all these transitions in Figure~\ref{fig:summary}.  

\section{Open Questions}

This is a fast-moving field, with a lively interplay of ideas between physics, probability, and computer science.  Here are some problems that remain open which the reader might enjoy.
\begin{itemize}
\item Do the information-theoretic and computational thresholds coincide for $q=3$, and $q=4$ in the assortative case?  For regular trees the Kesten-Stigum bound on tree reconstruction is tight for $q=3$ when the degree is sufficiently large~\cite{sly-reconstruction}.  Can this be extended to Galton-Watson trees where the number of children is Poisson-distributed?

\item Can we prove that reconstruction is hard for specific algorithms in the hard regime?  For instance, that natural Markov chains such as Glauber dynamics~\cite{LevinPeresWilmer2006,moore-mertens-book} have exponential mixing times, or that belief propagation with random initial messages succeeds with exponentially small probability?

\item In the hard regime, we can push an algorithm towards the accurate fixed point by giving it the true labels of a fraction of vertices.  This is known as semisupervised learning or side information, and has been studied both in physics~\cite{semisupervised} and computer science~\cite{kanade-mossel-schramm,mossel-xu-side-information}.  What can we say about phase transitions in this setting, as a function of the community structure and the amount or accuracy of the side information?

\item One interesting class of generalizations of the block model is the mixed-membership block model~\cite{ABFX08}, which allows communities to overlap: each vertex is associated with a $q$-dimensional vector, describing the extent to which it belongs to each community, and edges are generated based on these vectors.  Some versions of this model~\cite{ball-karrer-newman} can also be viewed as low-rank matrix factorization with Poisson noise.  Are there phase transitions in this model, parametrized by the density of the network and how much the communities overlap?

\pagebreak
\item Finally, the reader should be aware that community detection is just one of a vast number of problems in statistical inference.  Many such problems, including matrix factorization, sparse recovery, and others have phase transitions where the ground truth suddenly becomes impossible to find when the data becomes too noisy or incomplete, as well as hard regions where inference is information-theoretically possible but known polynomial-time algorithms such as Principal Component Analysis (PCA) provably fail; see e.g.~\cite{banks-isit}.  So if you enjoyed this article, you have your work cut out for you.
\end{itemize}

\section*{Acknowledgements}

This article is based on lectures and short courses I have given over the past few years, including at \'Ecole Normale Sup\'erieure, ICTS Bangalore, Northeastern, Institute d'\'Etudes Scientifiques de Cargese, \'Ecole de Physique des Houches, Beg Rohu, and the Santa Fe Institute, and seminars given at Harvard, Princeton, MIT, Stanford, Caltech, Michigan, Rutgers, Rochester, Northwestern, Chicago, Indiana, Microsoft Research Cambridge, l'Institut Henri Poincar\'e, the Simons Institute, and the Newton Institute.  I am grateful to these institutions for their hospitality, and to Emmanuel Abbe, V. Arvind, Caterina De Bacco, Josh Grochow, David Kempe, Florent Krzakala, Dan Larremore, Stephan Mertens, Andrea Montanari, Elchanan Mossel, Joe Neeman, Mark Newman, Guilhem Semerjian, Cosma Shalizi, Jiaming Xu, Lenka Zdeborov{\'a}, and Pan Zhang for collaborations, conversations, and/or comments on an earlier draft.  My work was supported by the John Templeton Foundation and the Army Research Office under grant W911NF-12-R-0012.

Vive la r\'esistance.

\bibliographystyle{plain}
\bibliography{refs}

\end{document}